\documentclass[aps,prd,superscriptaddress,nofootinbib,tighten,preprint]{revtex4}
\include{header}

\usepackage{amsfonts} 
\usepackage{amssymb}
\usepackage{amsmath}
\usepackage{graphicx}
\usepackage{hyperref}
\usepackage{placeins}
\usepackage{gensymb}
\usepackage{color}
\usepackage{times}
\usepackage{url,hyperref}
\usepackage{epstopdf} 
\usepackage[tight,footnotesize]{subfigure}
\def\slc#1{\setbox0=\hbox{$#1$}           
    \dimen0=\wd0                                 
    \setbox1=\hbox{/} \dimen1=\wd1               
    \ifdim\dimen0>\dimen1                        
       \rlap{\hbox to \dimen0{\hfil/\hfil}}      
       #1                                        
    \else                                        
       \rlap{\hbox to \dimen1{\hfil$#1$\hfil}}   
       /                                         
    \fi}

\begin{document}

\title{Prospects of Indirect Searches for Dark Matter at INO}

\author{Sandhya Choubey}
\email{sandhya@hri.res.in}
\affiliation{Harish-Chandra Research Institute (HBNI), Chhatnag Road, Jhunsi, Allahabad 211 019, India}
\affiliation{Department of Physics, School of
Engineering Sciences, KTH Royal Institute of Technology, AlbaNova
University Center, 106 91 Stockholm, Sweden}

\author{Anushree Ghosh}
\email{anushree.ghosh@usm.cl}
\affiliation{Universidad Tecnica Federico Santa Maria - Departamento de Fisica Casilla 110-V, Avda. Espana 1680, Valparaiso, Chile}

\author{Deepak Tiwari}
\email{deepaktiwari@hri.res.in}
\affiliation{Harish-Chandra Research Institute (HBNI), Chhatnag Road, Jhunsi, Allahabad 211 019, India}

\begin{abstract}
The annihilation of Weakly Interactive Massive Particles (WIMP) in the centre of the sun could give rise to neutrino fluxes. We study the prospects of searching for these neutrinos at the upcoming Iron CALorimeter (ICAL) detector to be housed at the India-based Neutrino Observatory (INO). We perform ICAL simulations to obtain the detector efficiencies and resolutions in order to simulate muon events in ICAL due to neutrinos coming from annihilation of WIMP in the mass range $m_\chi = (3-100)$ GeV. The atmospheric neutrinos pose a major background for these indirect detection studies and can be reduced using the fact that the signal comes only from the direction of the sun. For a given WIMP mass, we find the opening angle $\theta_{90}$ such that 90 \% of the signal events are contained within this angle and use this cone-cut criteria to reduce the atmospheric neutrino background. The reduced background is then weighted by the solar exposure function at INO to obtain the final background spectrum for a given WIMP mass. We perform a $\chi^2$ analysis and present expected exclusion regions in the $\sigma_{SD}-m_\chi$ and $\sigma_{SI}-m_\chi$, where $\sigma_{SD}$ and $\sigma_{SI}$ are the WIMP-nucleon Spin-Dependent (SD) and Spin-Independent (SI) scattering cross-section, respectively. For a 10 years exposure and $m_\chi=25$ GeV, the expected 90 \% C.L. exclusion limit is found to be $\sigma_{SD} < 6.87\times 10^{-41}$ cm$^2$ and $\sigma_{SI}  < 7.75\times 10^{-43}$ cm$^2$ for the  $\tau^{+} \tau^{-}$ annihilation channel and $\sigma_{SD} < 1.14\times 10^{-39}$ cm$^2$ and $\sigma_{SI} < 1.30\times 10^{-41}$ cm$^2$ for the $b~\bar b $ channel, assuming 100 \% branching ratio for each of the WIMP annihilation channel. 
\end{abstract} 

\maketitle

\section{Introduction}

Various cosmological and astrophysical observations strongly support the existence of the Dark Matter (DM) with an abundance of $\sim$27\%. The existence of DM was first postulated by Fritz Zwicky \cite{Zwicky:1933gu} who in an attempt to explain the dynamics of galaxies in the Coma galaxy cluster concluded that the most of the mass in the cluster must be invisible. Missing mass of spiral galaxies were also reported by Jan Oort and later confirmed by Vera Rubin (see \cite{Sofue:2000jx} for a more recent discussion) from the observation of flat galactic rotation curves which demanded that most of the matter of the galaxy is non-luminous and existed in the form of dark haloes. This inference has been supported by the weak \cite{Bartelmann:1999yn,vanUitert:2012bj} and strong \cite{Moustakas:2002iz} gravitational lensing data. Observations of the bullet cluster by the Chandra satellite \cite{Harvey:2015hha} further reinforced the idea of existence of a dark non-baryonic component of matter in the universe. Finally, the most precise measurement of the dark matter abundance of the universe comes from the measurements of the anisotropies of the cosmic microwave background spectra \cite{Hinshaw:2012aka}. The most recent analysis of the data from the Planck  satellite  \cite{Ade:2015xua} gives a dark matter abundance of $0.1172 \leq \Omega_{\rm DM}h^2 \leq 0.1226$ at 67 \% C.L.. \\

While the evidence for the existence of dark matter via its gravitational interactions is pretty strong, its particle nature remains largely unknown. Among the various possible candidates proposed, WIMP seem to be the most promising ones with masses ranging from a few GeVs to a few TeVs \cite{Jungman:1995df,Bertone:2004pz}. As the solar system moves through the DM halo, WIMP scatter off the nuclei in the celestial bodies like the sun and the earth. The scattered WIMP lose energy and could get gravitationally trapped by the gravitational potential of body and gradually sink to their cores. As WIMP annihilation rate scales with the square of its density, the core of these celestial bodies are the centres where WIMP could undergo annihilation, through various channels, into Standard Model (SM) particle-antiparticle pairs.  
The subsequent showering of these would give neutrinos whose energy spectra depend on the WIMP mass and annihilation channel\footnote{Models of extra-dimensions predict WIMP annihilating directly into $\nu_{e}\bar\nu_{e},\nu_{\mu}\bar\nu_{\mu}, \nu_{\tau}\bar\nu_{\tau}$ \cite{Appelquist:2000nn,Blennow:2009ag}.}. The neutrinos thus produced deep inside the sun, will undergo oscillations, interactions and regeneration as they propagate out of the core. Detection of these neutrinos, then, in principle, would provide information about the nature of DM {\it viz.}, its branching ratio, mass and cross-section. The signal neutrinos, on reaching the detector, interact with the medium and produce corresponding leptons. Various neutrino detectors like IceCube\cite{Aartsen:2016zhm}, Super-Kamiokande \cite{Tanaka:2011uf,Choi:2015ara} and ANTARES \cite{Albert:2016dsy} have been looking for such signatures and have put limits on the neutrino fluxes from annihilation of WIMP masses ranging from a few GeVs to a few TeVs. Prospects of indirect search of dark matter with these detectors have also been investigated in \cite{Kappl:2011kz,Das:2011yr,Rott:2011fh,Fornengo:2017lax}. \\

There is a proposal to build a 50 kt magnetised Iron CALorimeter (ICAL) detector at the India-based Neutrino Observatory (INO) \cite{Kumar:2017sdq}. The main physics goal of this detector would be to use atmospheric neutrino events to determine the neutrino mass hierarchy \cite{Ghosh:2012px,Devi:2014yaa,Ajmi:2015uda} and atmospheric neutrino parameters \cite{Thakore:2013xqa,Mohan:2016gxm,Kaur:2014rxa,Kaur:2017dpd} with good precision. However, it has been shown that one could use this detector to obtain competitive sensitivity to new physics scenarios such as sterile neutrinos \cite{Behera:2016kwr}, CPT violation in neutrinos \cite{Chatterjee:2014oda}, non-standard neutrino interactions \cite{Choubey:2015xha}, and magnetic monopoles \cite{Dash:2014fba} 
and decaying dark matter \cite{Dash:2014sza}, among others. Since the neutrinos produced in WIMP annihilations are in the energy range of a few GeV to 100 GeV for WIMP masses in the few GeV to 100 GeV range, ICAL should be able to efficiently detect these neutrinos and constrain the WIMP paradigm of dark matter. Prospects of indirect searches for dark matter at magnetised iron calorimeters were studied before in \cite{Mena:2007ty,Agarwalla:2011yy,Blennow:2013pya}.  All these studies were done assuming ad-hoc values for the detector specification such as detector efficiencies and energy and zenith angle reconstruction of the neutrino from WIMP annihilations. The atmospheric background suppression used in these earlier works were also ad hoc with the atmospheric neutrino background suppressed by a constant normalisation factor and without using the solar exposure function. In this work we perform a full simulation of the events from WIMP annihilations including reconstruction and charge identification efficiencies and energy and zenith angle resolutions obtained from a full detector simulations using a Geant4-based \cite{Agostinelli:2002hh,Allison:2006ve,Allison:2016lfl} code for ICAL \cite{Bhattacharya:2011qxc,Bhattacharya:2014tha}. We also carry out a detailed background suppression study in order to reduce the atmospheric neutrino events in the simulated data, which pose a serious background to indirect detection of WIMP in the sun.  We perform a $\chi^2$ analysis and present exclusion C.L. contours in the WIMP mass - WIMP-nucleon scattering cross-section plane. This work is a part of the on-going effort by the INO collaboration to study the physics reach of the ICAL detector. \\


 The paper is organised as follows. In Section \ref{sec:capture} we calculate the signal neutrino spectra due to WIMP annihilation in the sun. In Section \ref{sec:event} we describe the detector and the event generation procedure. Thereafter, in Section \ref{sec:background}, we describe the atmospheric background suppression scheme. In Section \ref{sec:analysis} we describe our statistical analysis. We present our main results in Section \ref{sec:results} and finally conclude in Section \ref{sec:summary}. In Appendix \ref{sec:append1} we describe our results on ICAL simulations to find the detector efficiencies and resolutions. 

\section{Signal neutrinos from WIMP annihilation}
\label{sec:capture}
The number of WIMP inside the sun as a function of time is given by the following differential equation \cite{Jungman:1995df},
\begin{equation}\label{eq:1}
 \frac{dN }{ d t}= C-C_{A} N^2-C_E N
 \,,
\end{equation}
where the three terms on the right-hand side correspond to capture of WIMP inside the sun, annihilation inside the sun's core and evaporation from its surface, respectively. 
The effect of evaporation from the sun is seen to be important only for very light WIMP \cite{Spergel:1984re,Gould:1987ju,Krauss:1985aaa,Griest:1986yu}. Since we will be working with WIMP masses above 3 GeV, we will neglect the last term of Eq.~(\ref{eq:1}) in this paper. Since each annihilation reduces the number of WIMP by two units, the rate of depletion of WIMP ($\Gamma_{A}$) is twice the annihilation rate in the sun and hence 
\begin{equation} 
\Gamma_{A}=  \frac{1}{2} C_{A}N^2
\,.
\end{equation}
Solving Eq.~(\ref{eq:1}) for $N$, we find the annihilation rate at any given time as 
\begin{equation} 
    \Gamma_{A}=\frac{1}{2}C\tanh{^2}(t/\tau) 
    \,,
\end{equation}
where $\tau=(CC_{A})^{-1/2}$ is the time required for equilibrium to be established between the capture and annihilation of WIMP in the sun. If the age of the sun is greater than the 
equilbrium time scale  ($t_{\odot} = 1.5 \times10^{17}$ s $>> \tau $), then $\Gamma_{A}= \frac{1}{2}C$. \\

The WIMP capture rate due to Spin Independent (SI) interactions in the sun is given by \cite{Gould:1991hx,Jungman:1995df},
\begin{equation} 
\label{eq:2}
 C_{SI} = c \left(\frac{1~\textup{GeV}}{m_\chi}\right) 
  \left( \frac{\rho_{local}}{0.3~ \textup{GeV}/\textup{cm}^{3}}  \right)  \left(\frac{270~ \textup{km}/\textup{s}}{\bar v_{\textup{local}}} \right) \sum_{i} F_{i}(m_{\chi}) \sigma_{SI}^i f_{i}\phi_{i}\frac{S(m_{\chi}/m_{Ni})}{m_{Ni}/(1~{\rm GeV})}
 \,,
\end{equation}
where $c=4.8\times 10^{24}$ s$^{-1}$, $m_\chi$ is the mass of DM, $\rho_{local}$ is the local DM density and $\bar v_{\textup{local}}$ is the DM velocity dispersion in the halo, the summation has to be carried out over all the nuclei in the sun,  $F_i(m_\chi$) is the form-factor suppression for the capture of a WIMP of mass $m_\chi$ with the $i^{th}$ nuclei, $m_{N_i}$ is the mass (in GeV) of the $i^{th}$ nuclear species, $f_{i}$ is the mass fraction of the $i^{th}$ element, and $\sigma_{SI}^i$ is the cross-section for elastic scattering of the WIMP from the $i^{th}$ nucleus via SI interaction in units of $10^{-40}$ cm$^2$, $\phi_i$ gives the distribution of the $i^{th}$ element in the sun, while $S(m_\chi / m_{N_i})$ is the kinematic suppression factor for the capture of the WIMP. $\sigma^{i}_{SI}$ can be related to  $\sigma_{SI}$ as $\sigma^{i}_{SI} = \sigma_{SI} A^{2}_{i} \bigg( \frac{\mu_{\chi N_{i}} }{\mu_{\chi p}} \bigg)^{2}$; where $A$ is the atomic number, $\mu$ is the reduced mass and in approximation $m_{N_{i}} =A_{i}m_{p} $, where $m_{p} $ is the proton's mass.\\

\begin{figure}[tbp]
\centering 
\includegraphics[width=.75\textwidth,origin=c]{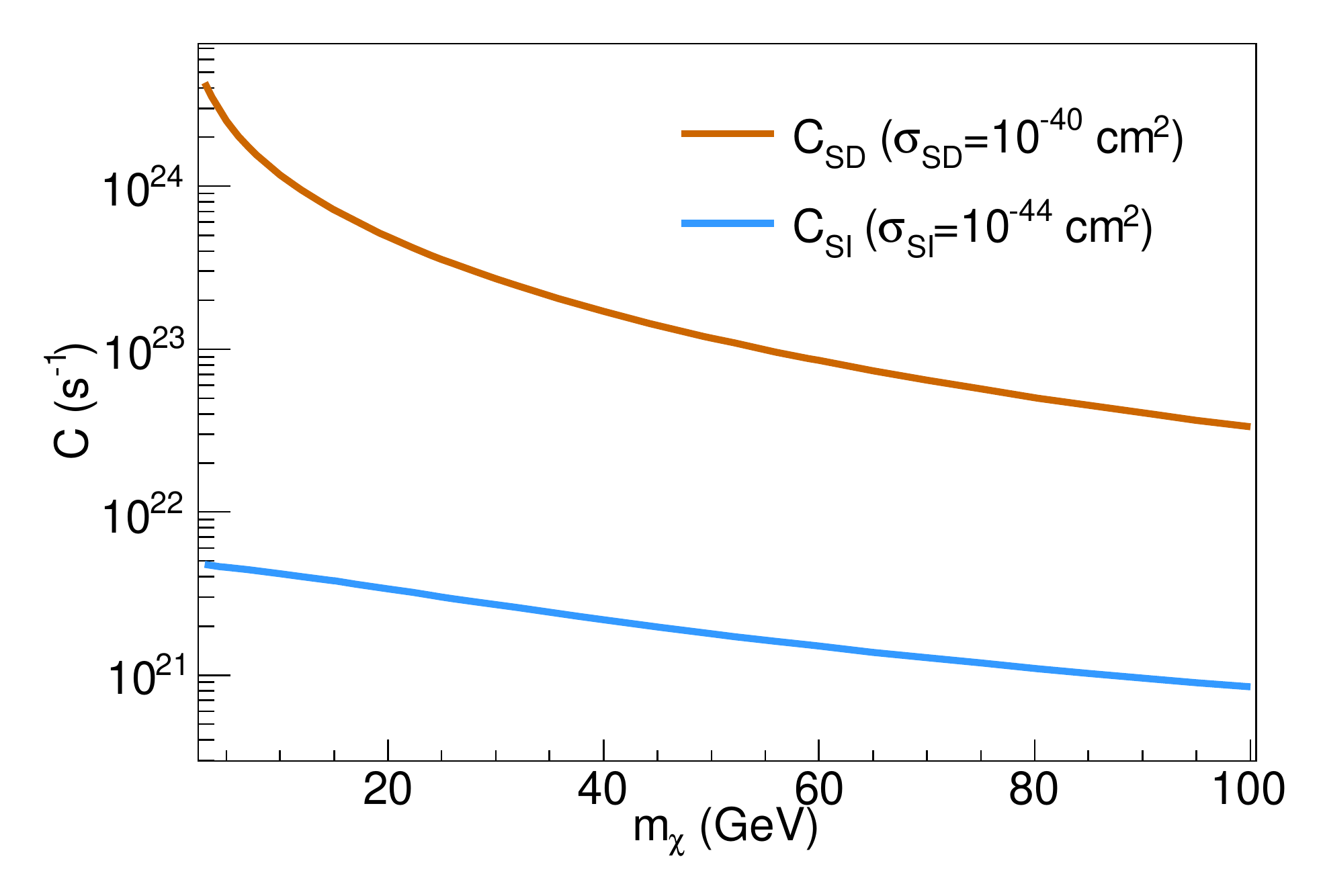}
\caption{\label{fig:1} SI and SD capture rates for WIMP in the core of the sun as a function of WIMP mass $m_{\chi}$. }
\end{figure}

The Spin Dependent (SD) capture rate of WIMP due to elastic scattering off sun's nuclei is given by \cite{Mena:2007ty},
\begin{equation} 
 C_{SD}= 9\times 10^{24}~\textup{s}^{-1} \left(\frac{\sigma_{SD}}{10^{-2}~\textup{pb}}\right) \left(\frac{50~\textup{GeV}}{m_\chi}\right)^2 
  \left( \frac{\rho_{local}}{0.3~ \textup{GeV}/\textup{cm}^{3}}  \right) \left(\frac{270~ \textup{km}/\textup{s}}{\bar v_{\textup{local}}} \right)
  \,,
\end{equation}
where as before, $m_\chi$ is the DM mass, $\rho_{\textup{local}}$ is the local DM density, $v_{\textup{local}}$ is the DM velocity dispersion in the halo and $\sigma_{SD}$ is the SD WIMP-nucleon cross- section. 
Figure~\ref{fig:1} shows the SD and SI capture rates for the sun for an assumed WIMP-nucleon cross-section for each case. \\

The WIMP annihilate into pairs of standard model leptons, quarks, gauge and Higgs bosons. Their subsequent hadronisation and/or decay give rise to neutrinos. We will also consider the situation where the WIMP annihilate directly into neutrino-antineutrino pairs. The differential neutrino flux at the detector coming from WIMP annihilations in the sun is given by
\begin{equation} 
 \frac{dN_{\nu}'}{d\Omega dt dE_{\nu}} = \frac{\Gamma_{A}}{4\pi R^{2}} \sum_{j=1} {\rm BR}_j \frac{dN_{j}}{dE_{\nu}}
 \,,
 \label{diffflux}
\end{equation}
where $\Gamma_A$ is defined above in terms of $C_{SI}$ or $C_{SD}$, $R$ is the distance traveled by the neutrinos and $dN_j/dE_\nu$ is the differential flux for a given annihilation channel $j$, where $j$ could be $W^+W^-$, $b\bar b$, $c\bar c$, $\tau^{+}\tau^{-}$ and so on. The sum in Eq.~(\ref{diffflux}) is over all possible channels $j$ and the sum has to be weighted with the branching ratio ($BR_{j}$) of the particular channel $j$. These branching ratios can be calculated within the framework of specific models. Since we consider a generic WIMP scenario, we will take only one annihilation channel at a time and assume 100\% branching ratio for that channel. The expected sensitivity limit calculated for each channel for 100 \% branching ratio indicates the limit expected for that particular channel alone. A given WIMP model will predict a mixture of these channels with varying branching ratios, and hence the corresponding sensitivity limit will lie somewhere in the region bounded by the best and worst limit expected from these various channels. For instance, the expected sensitivity limit for a DM model that predicts annihilation of WIMPs into $\tau^{+}\tau^{-}$ with 20 \% BR and into  $b\bar b$ with 80 \% BR, will lie in a region between the expected sensitivity for these two channels with 100\% BR, and closer to the latter channel. It should be noted, however, that this example model explicitly assumes only two annihilation channels. In case WIMP annihilate to less competitive annihilation channels such as $d \bar{d}$, $u \bar{u}$, $s \bar{s}$ etc the sensitivity limits would decrease proportionately. Similarly, WIMP annihilation to channels like $e^{+}$ $e^{-}$ etc which do not contribute to neutrino fluxes, would weaken the sensitivity. 
The annihilation of WIMP can occur through various channels : $c\bar c$, $b\bar b$, $t\bar t$, $e^{+}e^{-}$, $\mu^{+} \mu^{-}$, $\tau^{+}\tau^{-}$, $W^+W^-$, $Z^0Z^0$, $g~g$, $d \bar{d}$, $u \bar{u}$, $s \bar{s}$ etc. Among these channels, electrons are stable and the muons would interact and get absorbed inside the sun before they could produce high energy neutrinos. Hence these are not relevant to our analysis. Annihilation of DM into particles like protons, anti-deutrons, gamma rays will also not produce neutrino fluxes and hence not considered in our work. The quark-antiquark annihilation channels like $u \bar u$, $d \bar d$ and $s \bar s$ will produce a weaker neutrino spectra, in fact a few order of magnitude smaller than the $g~g$ annihilation channel and hence we do not consider them in our analysis. The $g~g$ annihilation channel is not competitive, but we quote the sensitivity limits for this channel. We will work with WIMP masses between a few GeVs to upto 100 GeV, and hence the channels $W^+W^-$, $Z^0Z^0$, $t\bar t$ are not kinematically relevant as they open up from masses 80.4 GeV, 91.2 GeV and 173 GeV, respectively. The annihilation to Higgs
boson has also not been considered for the same reason. In WIMP models like Kaluza-Klein (KK) dark matter, WIMP could directly undergo annihilation into neutrinos $\nu_{e}\bar\nu_{e},\nu_{\mu}\bar\nu_{\mu}, \nu_{\tau}\bar\nu_{\tau}$.
We have considered these channels in our analysis.\\\\
The spectra of neutrino fluxes due to WIMP annihilation in the sun have been calculated in detail in \cite{Cirelli:2005gh, Blennow:2007tw}. In this work, for simulating the WIMP annihilations into standard model particle-antiparticle pairs in the centre of the sun, the subsequent propagation of the daughter particles and finally the neutrinos, we use WIMPSIM \cite{Edsjo:2007xyz} package. WIMPSIM uses Nusigma \cite{Edsjo:2007abc} for simulation of neutrino-nucleon interactions and PYTHIA \cite{Sjostrand:2006za} for the hadronisation, decay and production of neutrinos. The upper WIMP mass considered for all the annihilation channels is 100 GeV and the lower mass considered is 3 GeV with the exception of $b\bar b$ whose lower mass limit has been taken to be 7 GeV. The propagation of neutrinos includes full three flavor neutrino oscillations with the oscillation parameters given in Table~\ref{tab:1}. We consider normal mass hierarchy for the neutrinos in our analysis.\\

\begin{table}[tbp]
\centering
\begin{tabular}{|lr|c|}
\hline
Paramter & Best-Fit Value\\
\hline 
$\theta_{12}$ & $34^{\circ}$ \\
$\theta_{13}$ & $9.2^{\circ}$ \\
$\theta_{23}$ & $45^{\circ}$ \\
$\delta$ & $0$ \\
$\Delta m^2_{21}$ & $7.5\times 10^{-5} \textup{eV}^2$ \\
$\Delta m^2_{31}$ & $2.4\times 10^{-3} \textup{eV}^2$\\
\hline
\end{tabular}
\caption{\label{tab:1} Oscillation parameters used in the simulations.}                                                                                                                                                
\end{table}
\begin{figure}[tbp]
\centering 
\includegraphics[width=.75\textwidth,]{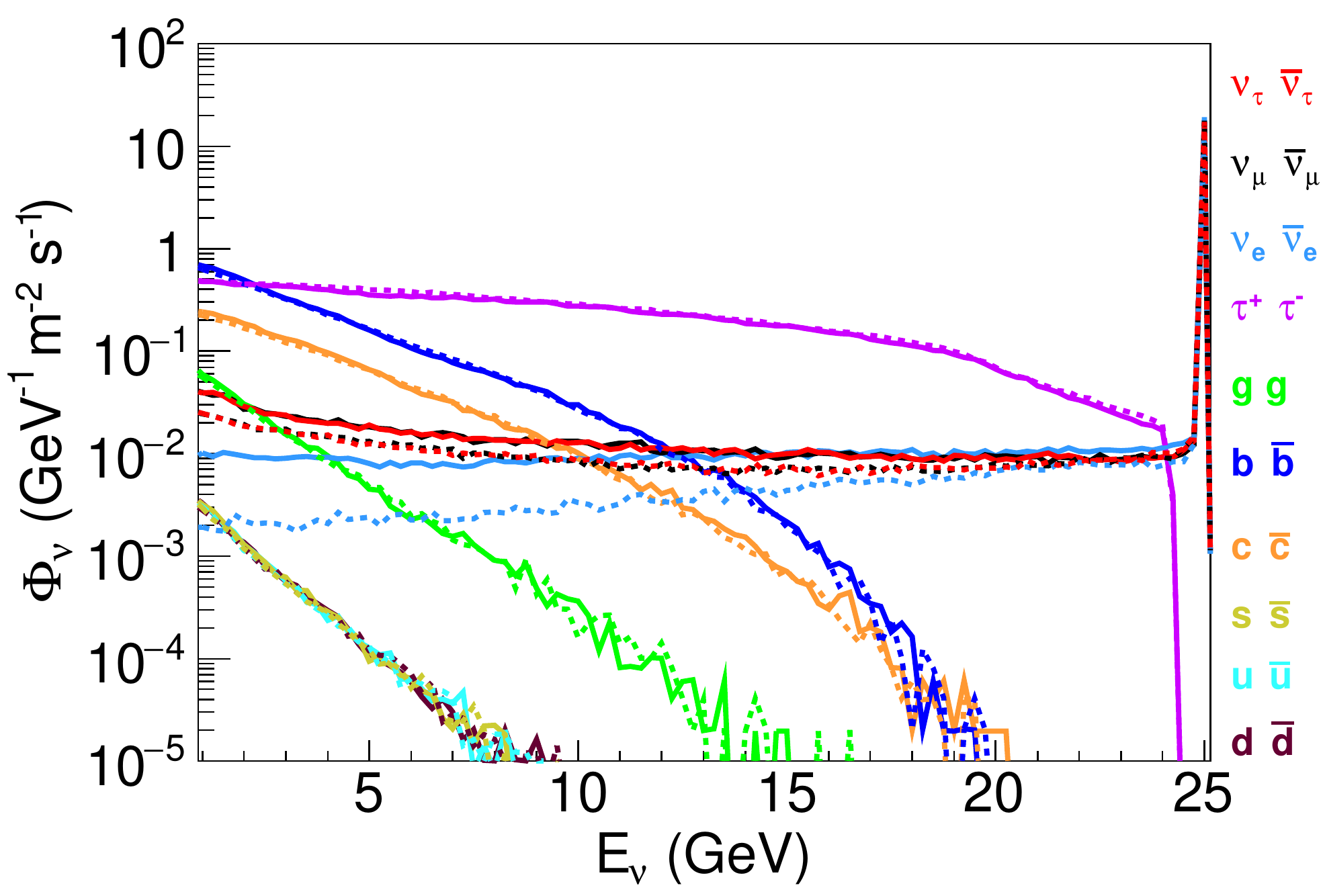}
\caption{\label{fig:2} The $\nu$ (solid) and $\bar{\nu}$ (dashed) fluxes at ICAL due to annihilation of 25 GeV WIMPs in the sun for channels for various annihilation channels with $\sigma_{SD}=10^{-39}$ cm$^{2}$. The fluxes due to SI interactions in the sun are similar and scaled by the appropriate $\sigma_{SI}$. }
\end{figure}

The neutrino and antineutrino fluxes (in units of GeV$^{-1}$m$^{-2 } \Omega^{-1 }$s$^{-1}$) at ICAL due to WIMP annihilations in the sun is shown in Fig.~\ref{fig:2}. We use the following values: $ \rho_{\textup{local}} =0.3~ \textup{GeV}/ \textup{cm}^3, v_{\textup{local}}= 270 ~\textup{km}~ \textup{sec}^{-1}$ and 100 \% BR for each of the channels. We show the fluxes for WIMP mass of 25 GeV and WIMP-nucleon scattering cross-sections of $\sigma_{SD}=10^{-39}$ cm$^{2}$. Note from Fig.~\ref{fig:2} that for WIMP mass of 25 GeV, the fluxes for both neutrinos as well as antineutrinos are nearly same. 
The above feature is seen to be true for nearly all WIMP masses. Also note that for the $\tau^{+} \tau^{-}$ channel, the neutrino fluxes fall by only about 1 order of magnitude between neutrino energies 0 and 25 GeV, which is the maximum possible neutrino energy from a 25 GeV WIMP. For the $b \bar{b}$ channel on the other hand, the fluxes are lower than those for the $\tau^{+} \tau^{-}$ channel to begin with, and subsequently fall sharply by many orders of magnitude by $E_\nu = 15$ GeV. For higher energies, the neutrino flux from the $b \bar{b}$ channel is relatively negligible. The fluxes due to $c \bar{c}$ channel are lower than $b \bar{b}$ for all neutrino energies but follow a similar trend. The contribution from $g g$ channel is an order less than $b \bar{b}$. The fluxes due to $u \bar u$, $d \bar d$ and $s \bar s$ channels are pretty indistinguishable from each other and are several orders of magnitude lower than $b \bar{b}$ making them least competitive. The KK channels give rise to monoenergetic neutrino fluxes at high energies and hence will give the most sensitive limits. Therefore, it is evident that the indirect detection  bounds from observation of these neutrinos and antineutrinos will be stronger when one considers the $\nu \bar{\nu}$ or $\tau^{+} \tau^{-}$ channels compared to when one takes the softer channels such as $b \bar{b}$, $c \bar{c}$ etc. As mentioned above, we do not consider the other channels with weaker neutrino flux strength in our discussion on the expected sensitivity to indirect detection of dark matter in ICAL. Within the context of specific WIMP models, of course, the individual BR for each of the channels can be calculated and then one can find the expected sensitivity of ICAL to indirect detection within that given model. For illustration, however, we give the expected sensitivity limits for the $\tau^{+} \tau^{-}$, $b \bar{b}$, $c \bar{c}$ and $g g$ channels with 100 \% BR as mentioned above, and also consider WIMP annihilation channels $\nu_{e}\bar\nu_{e},\nu_{\mu}\bar\nu_{\mu}, \nu_{\tau}\bar\nu_{\tau}$ in our analysis and calculate the corresponding expected sensitivities. The Fig.~\ref{fig:2} is shown for benchmark values of WIMP mass and cross-sections. The fluxes for other values of $\sigma_{SD}$  simply scale with the value of the cross-section, however, for other values of WIMP masses the spectral shape changes. Nevertheless, the above mentioned features regarding the fluxes from different annihilation channels remain true for all WIMP masses. 
\\
\section{Event generation at ICAL}
\label{sec:event}

The ICAL detector is an upcoming 50 kt iron calorimeter detector at the proposed India-Based Neutrino Observatory in Theni district of Tamil Nadu, India. It will consist of 150 layers of glass RPCs (Resistive Plate Chambers) interspersed with iron plates. The details of the detector has been described in detail elsewhere \cite{Kumar:2017sdq}. The neutrino (or antineutrino) on entering the detector interacts with detector nucleon to produce a muon (or antimuon) and hadron(s). The muon (or antimuon) produces a clean track in the magnetised iron, with the muon and antimuon bending in opposite directions, giving the detector an excellent charge identification capability. The hadron(s) produces a shower, which can also be detected. The detector is expected to have good muon energy and angle resolution, reasonably good muon reconstruction efficiency and excellent charge identification efficiency. 
In this work, we explore the versatility of the detector for probing dark matter.  As we will see, the excellent angular resolution of muons could be used to put competitive limits on the WIMP scenario. \\

We calculate the signal and background muon and antimuon events using the event generator GENIE \cite{Andreopoulos:2009rq}, which has been suitably modified to include the ICAL geometry. The signal events are calculated using the neutrino fluxes as described in Section \ref{sec:capture}. The atmospheric neutrino background events are calculated using the Honda fluxes \cite{PhysRevD.83.123001} given for the INO site. These events are then passed through our reconstruction code whereby we apply detector energy and angle resolutions, as well as, reconstruction and charge identification efficiencies to get the final events. These muons are binned in reconstructed energy and zenith angle bins. The muon reconstruction efficiency, muon charge identification efficiency, muon zenith angle resolution and muon energy resolution values are obtained though the Geant4-based \cite{Agostinelli:2002hh} detector simulation code for ICAL, developed by the INO collaboration. The details of our simulation procedure and snapshots of some of the results are given in Appendix \ref{sec:append1}. The results in Appendix \ref{sec:append1} show that the energy and the angle resolutions for the muons are functions of both muon energy and muon zenith angle. Same is true for the charge identification efficiency and reconstruction efficiency. The detector efficiencies obtained as such and shown in Figs.~\ref{fig:16} and \ref{fig:17} are then implemented onto the signal and background events as follows:\begin{equation} 
 N'^{th} _{ij} =  \mathcal{N} \sum_{k} \sum_{l} K^{k} _{i} (E^{k} _{T}) M^{l} _{j} (\cos{\Theta^{l} _{T}}) \left( \varepsilon_{kl} \mathcal{C}_{kl} n_{kl} (\mu^{-}) + \bar{\varepsilon}_{kl} (1- \bar{\mathcal{C}}_{kl}) n_{kl} (\mu^{+}) \right) 
 \label{evtrue}
\end{equation}
where the indices $i$ and $j$ denote the measured energy and zenith angle bin of the muon, $\mathcal{N}$ is the normalisation corresponding to a specific exposure in ICAL, $E^k_{T}$ and $\cos{\Theta^l_{T}}$ are the true (kinetic) energy and true zenith angle of the muon, where the indices $k$ and $l$ denote the true energy and true zenith angle bin of the muon. 
The quantities $n_{kl}(\mu^{-})$ and $n_{kl}(\mu^{+})$ are the number of $\mu^{-}$ and $\mu^{+}$ events in the $k^{th}$ true energy and $l^{th}$ true zenith angle bin, respectively. The quantities $\varepsilon_{kl}$ and  $\bar{\varepsilon}_{kl}$ are the reconstruction efficiencies of $\mu^{-}$ and $\mu^{+}$ respectively for the $k^{th}$ energy and the $l^{th}$ zenith angle bin. $\mathcal{C}_{kl}$ and $\bar{\mathcal{C}}_{kl}$ are the corresponding charge identification quantities.
The muon energy and angle smearing are then implemented by folding in the Gaussian resolution functions
$K^{k}_{i}$ and $M^{l}_{j}$ in Eq.~(\ref{evtrue}). These are given as, 
\begin{equation} 
  K^{k} _{i}= \int _{E_{L_{i}}} ^{E_{H_{i}}} dE \frac {1} {\sqrt{2\pi}\sigma_{E}} \textup{exp} \left( -\frac {({E^{k}_{T}-E})^{2}} {2\sigma^{2}_{E}} \right)
\,,
\end{equation}
\begin{equation} 
 M^{l} _{j}(\cos\Theta_{T}^{l})  = \int _{\cos{\Theta_{L_{j}}}} ^{\cos{\Theta_{H_{j}}}} d(\cos{\Theta}) \frac {1} {\sqrt{2\pi}\sigma_{\cos_{\Theta}}} \textup{exp} \left( -\frac {({\cos{{\Theta}^{k}_{T}}-\cos{\Theta}})^{2}} {2\sigma^{2}_{\cos_{\Theta}}} \right)
 \,,
\end{equation}
and integrating out the true energy and angle of the muons, 
where $E$ and $\cos{\Theta}$ are the measured (kinetic) energy and zenith angle of the muon and 
the values of $\sigma_{E}$ and $\sigma_{\cos{\Theta}}$ are given in Figs.~\ref{fig:19} and \ref{fig:20}, respectively.
Similar expressions can be written for the $\mu^+$ events. Throughout this work we will work with simulated events for 10 years of running of ICAL.

\begin{figure}[tbp]
\centering
\includegraphics[width=.75\textwidth]{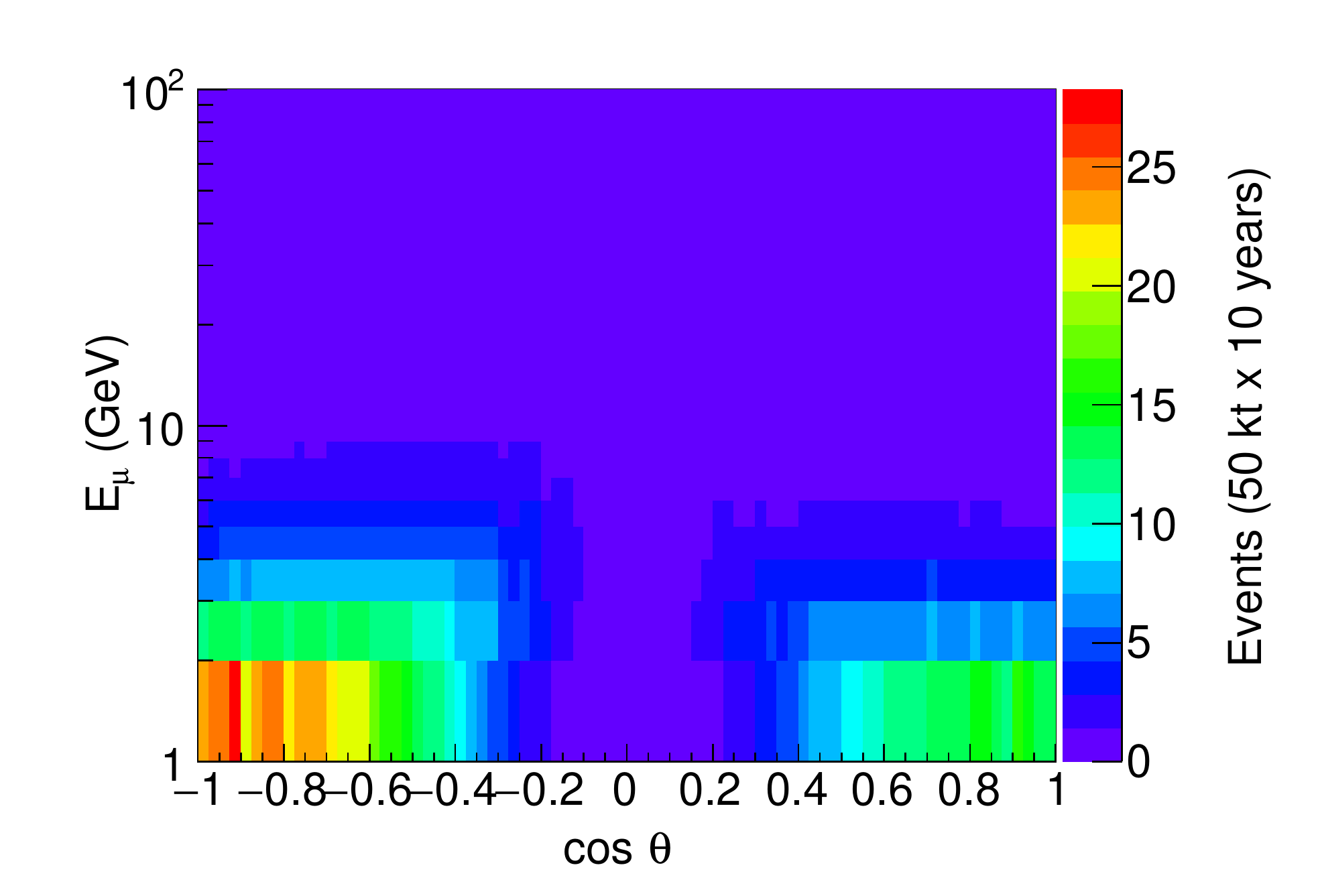}
\caption{\label{fig:3} The $\mu^{-}$ event distribution at ICAL due to atmospheric neutrino background for 50 $\times$ 10 kt-years of ICAL exposure. ICAL has zero efficiency for horizontal tracks which is reflected in the bins around $\cos\theta =0$. Note that $\cos\theta = 1$ represents upward going muons in ICAL convention.}
\end{figure}

\section{Atmospheric neutrino background suppression}
\label{sec:background}

The atmospheric neutrinos pose a large background\footnote{There is another source of background where high energy neutrinos are produced in reactions when cosmic rays hit the solar corona. However, this background is of order of a few neutrino events per year, and we ignore them for our present analysis \cite{Edsjo:2017kjk}.} to the indirect detection signal. Fig.~\ref{fig:3} shows the reconstructed $\mu^{-}$ event distribution due to atmospheric neutrino background. These neutrinos have energies similar to those of neutrinos from WIMP annihilation for WIMP masses 3-100 GeV and hence atmospheric neutrino events are indistinguishable from indirect detection events in ICAL. However, there are two features in which the signal neutrinos are different from the atmospheric neutrinos. Firstly, the signal flux, as we had seen in Section \ref{sec:capture}, has an energy dependence that is quite different from the energy dependence of the atmospheric neutrinos which falls sharply as $\sim E_\nu^{-2.7}$. Therefore, an analysis binned in energy should be able to discriminate between the two kinds of events. More importantly, unlike the neutrinos from WIMP annihilation which come from the direction of the sun, the atmospheric neutrinos have a distribution over all zenith and azimuth angular bins. We can exploit this feature for an effective background suppression. In what follows, we will describe in detail our cone-cut analysis method for suppressing the atmospheric neutrino background. \\

\begin{figure}[tbp]
\centering 
\includegraphics[trim=1 1 1 1,clip,width=.45\textwidth]{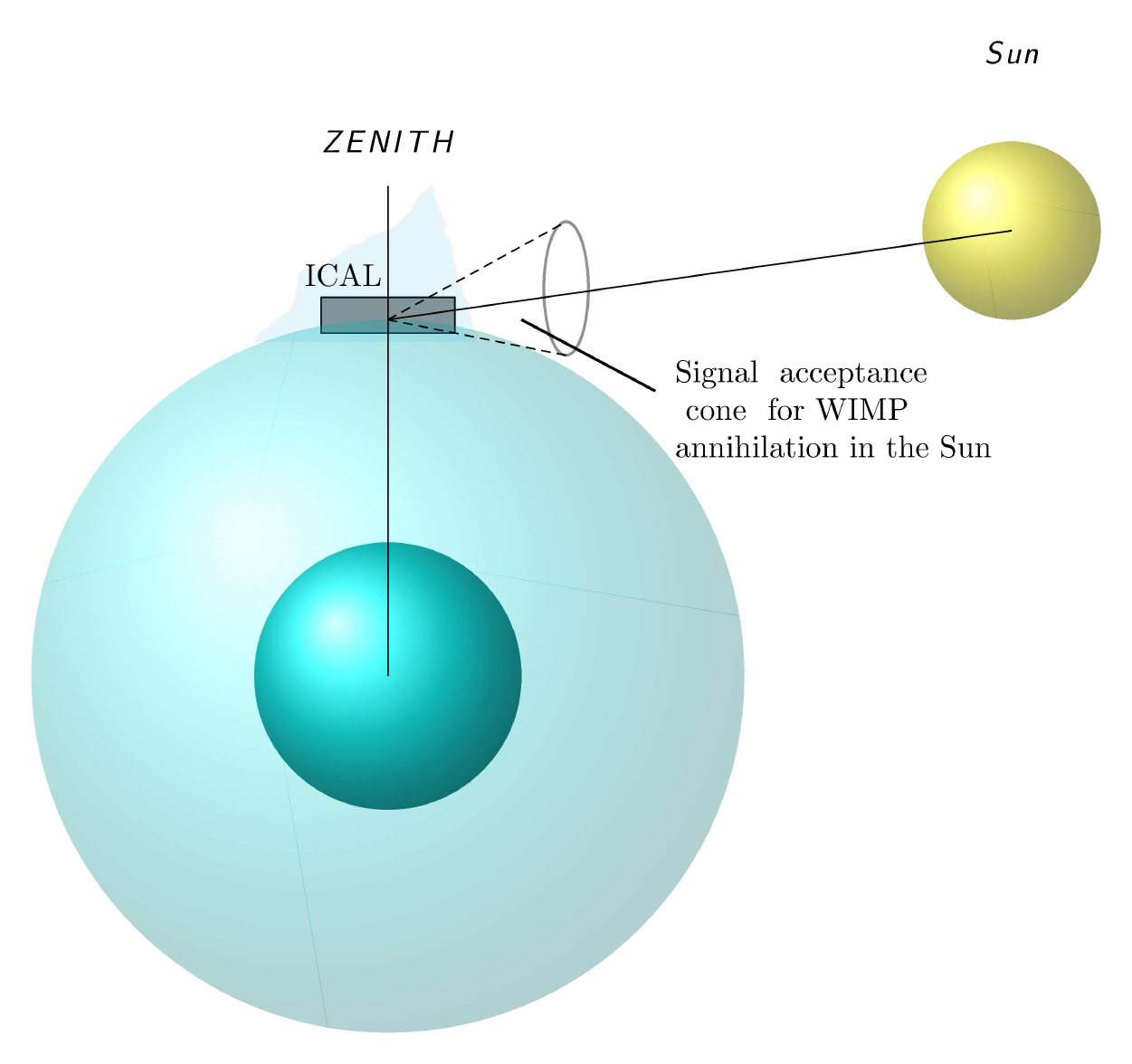} 
\caption{\label{fig:4}  The cone regions where signal from the WIMP annihilations are expected for the sun. 
}
\end{figure} 

Since our signal comes only from the sun while the atmospheric neutrino comes from all sides, we accept events only from the direction of the sun.  The signal neutrinos will be coming from the direction of the sun and the associated scattered lepton produced at the detector (muon in our case) will make an angle ($\theta_{\nu_\mu}$) with the parent neutrino. The angle $\theta_{\nu_\mu}$ depends only on the energy of the parent neutrino and the detector medium. Due to finite detector resolution there will be smearing effects and the reconstructed muon direction will be slightly different compared the true direction. However, since muon angle resolution is rather good for ICAL (cf. Appendix \ref{sec:append1}), we do not apply detector angular resolutions at this stage of background suppression for simplicity. That is to say, we calculate $\theta_{\nu_\mu}$ using the generator level information {\it i.e.}, using the true direction of neutrino and the corresponding muon rather than using true direction of neutrino and reconstructed muon direction. \\

\begin{figure}[tbp]
\centering 
\includegraphics[width=.45\textwidth,]{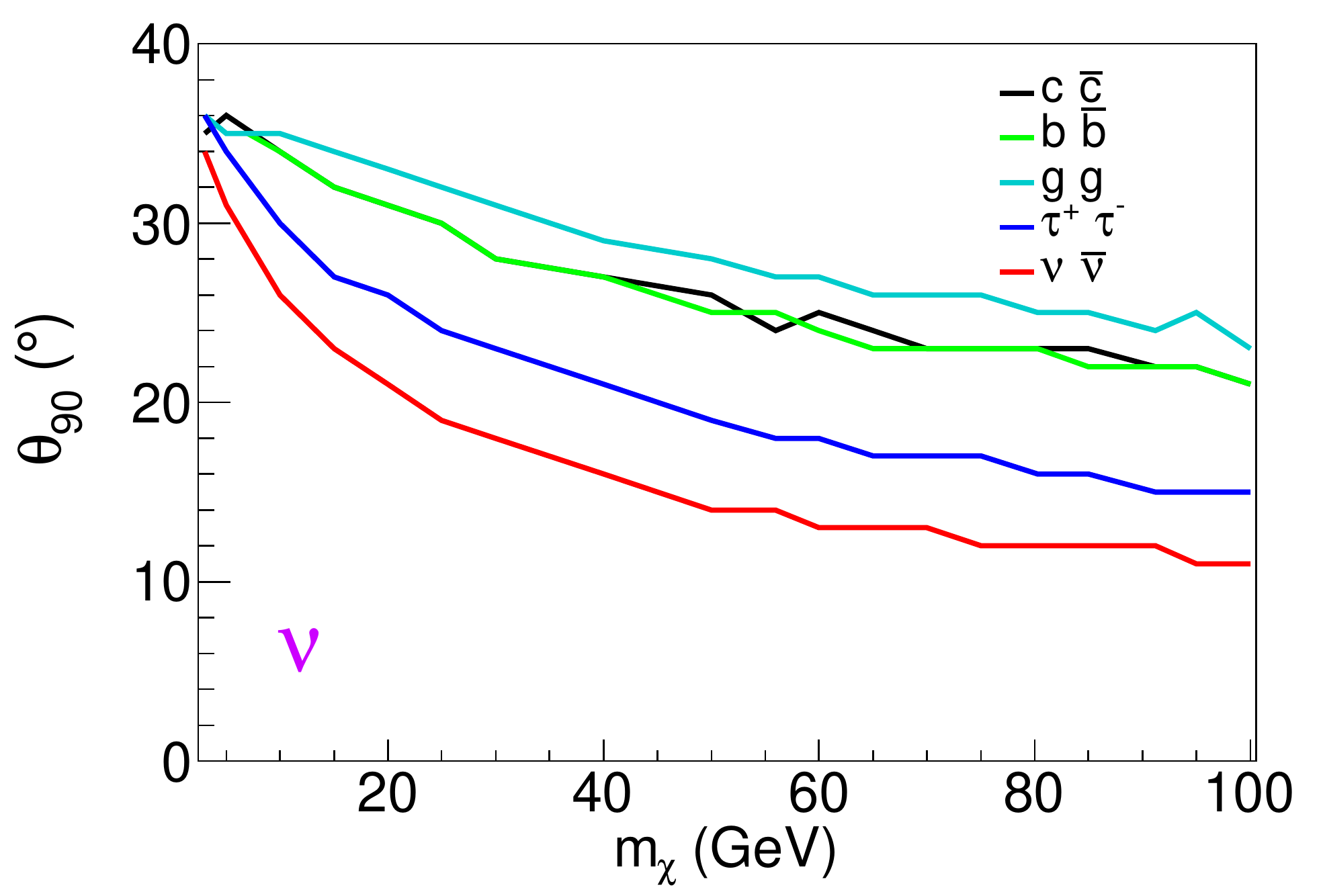}
\hfill
\includegraphics[width=.45\textwidth,origin=c]{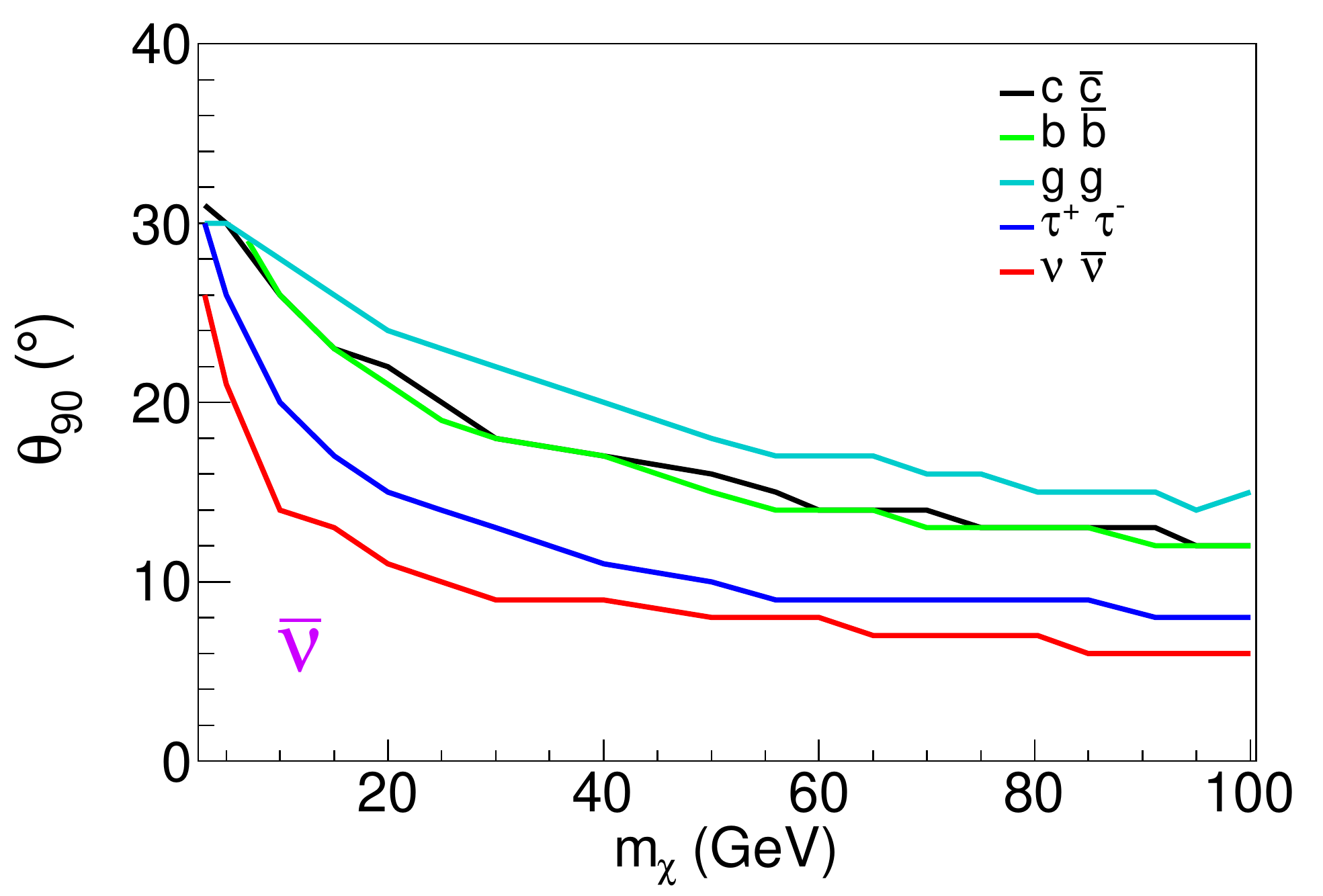}
\caption{\label{fig:5} 90 \% cone-cut values obtained for the sun for neutrinos (LEFT) and antineutrinos (RIGHT) as a function of the WIMP mass $m_\chi$. }
\end{figure}

We define cone angle $\theta_{90}$ as the half angle of the cone that contains 90 \% of the signal muons, the axis of the cone being in the direction of the parent neutrino. The cartoon showing our geometrical cone-cut criteria is given in Fig.~\ref{fig:4}. The higher energy neutrinos will have a narrower cone opening while lower energy ones will have a broader $\theta_{90}$. Since the neutrino energy is determined by the mass of the annihilating WIMP, we expect the neutrino flux from annihilation of heavier WIMP to produce more  muons peaked along the direction of the sun and hence have narrower cone angle $\theta_{90}$ than neutrino flux produced by lighter WIMP. For the same reason, $\theta_{90}$ for the $\tau^+\tau^-$ channel is expected to be smaller than $\theta_{90}$ for the $b\bar b$ channel. Likewise, among the channels considered, we expect the largest and smallest value of $\theta_{90}$ for $g g$ and $\nu \bar{\nu}$ \footnote{The fluxes, the cone-cut angles and hence the expected sensitivities due to $\nu_{e}\bar\nu_{e},\nu_{\mu} \bar\nu_{\mu} and \nu_{\tau} \bar\nu_{\tau}$ are almost indentical. Throughout this paper, unless otherwise specified by a subscript, we have taken $\nu_{\tau} \bar\nu_{\tau}$ as the representative of all neutrino flavors and indicated by $\nu \bar{\nu}$.} respectively. Using WIMPSIM and GENIE, we calculate $\theta_{90}$ for each WIMP mass and for a given annihilation channel in the sun. Fig.~\ref{fig:5} shows the $\theta_{90}$ calculated for the sun as a function of the WIMP mass ($m_{\chi}$). The different lines correspond to different annihilation channels. For each WIMP mass and annihilation channel, we place $\theta_{90}$ cone around the neutrino direction and accept events that fall within this cone. As expected the cone-cut angle $\theta_{90}$ is smaller for the $\tau^+\tau^-$ channel compared to the $b\bar b$ channel since the former is harder compared to the latter. The $\theta_{90}$ for the antineutrinos is seen to be smaller since for the same energy, the $\mu^+$ events from antineutrinos are seen to be more forward peaked compared to the $\mu^-$ events coming from neutrinos. \\

\begin{figure}[tbp]
\centering 
\includegraphics[width=.75\textwidth,]{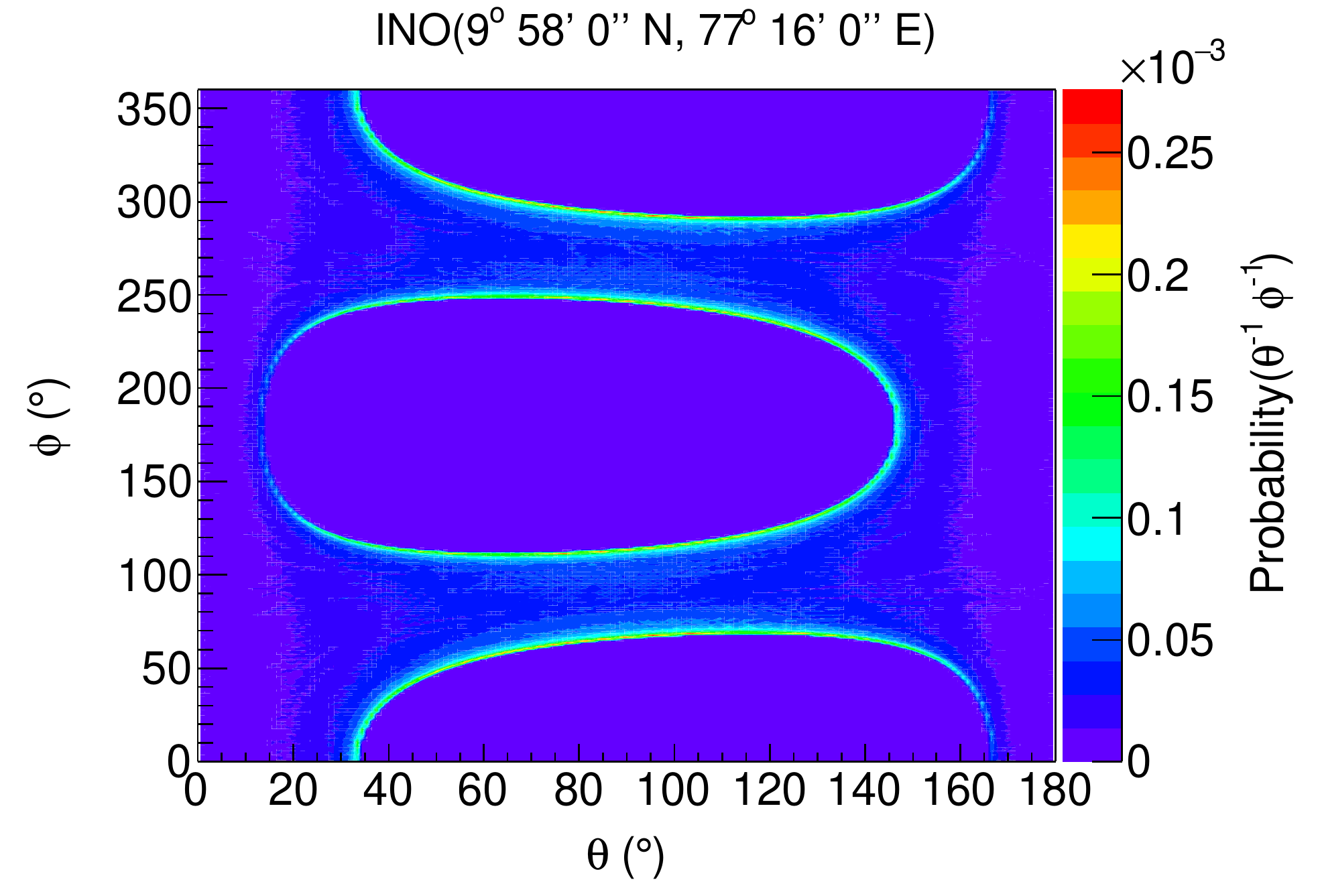}
\caption{\label{fig:6} The figure shows the probability of solar exposure for each zenith ($\theta$) and azimuthal ($\phi$) angle bin for INO's geographical coordinates. 
Here, $\theta = 180^{\circ}$ represents zenith. The plot has been obtained through SLALIB routine of WIMPSIM.}
\end{figure}
\begin{figure}[tbp]
\centering
\includegraphics[width=.75\textwidth]{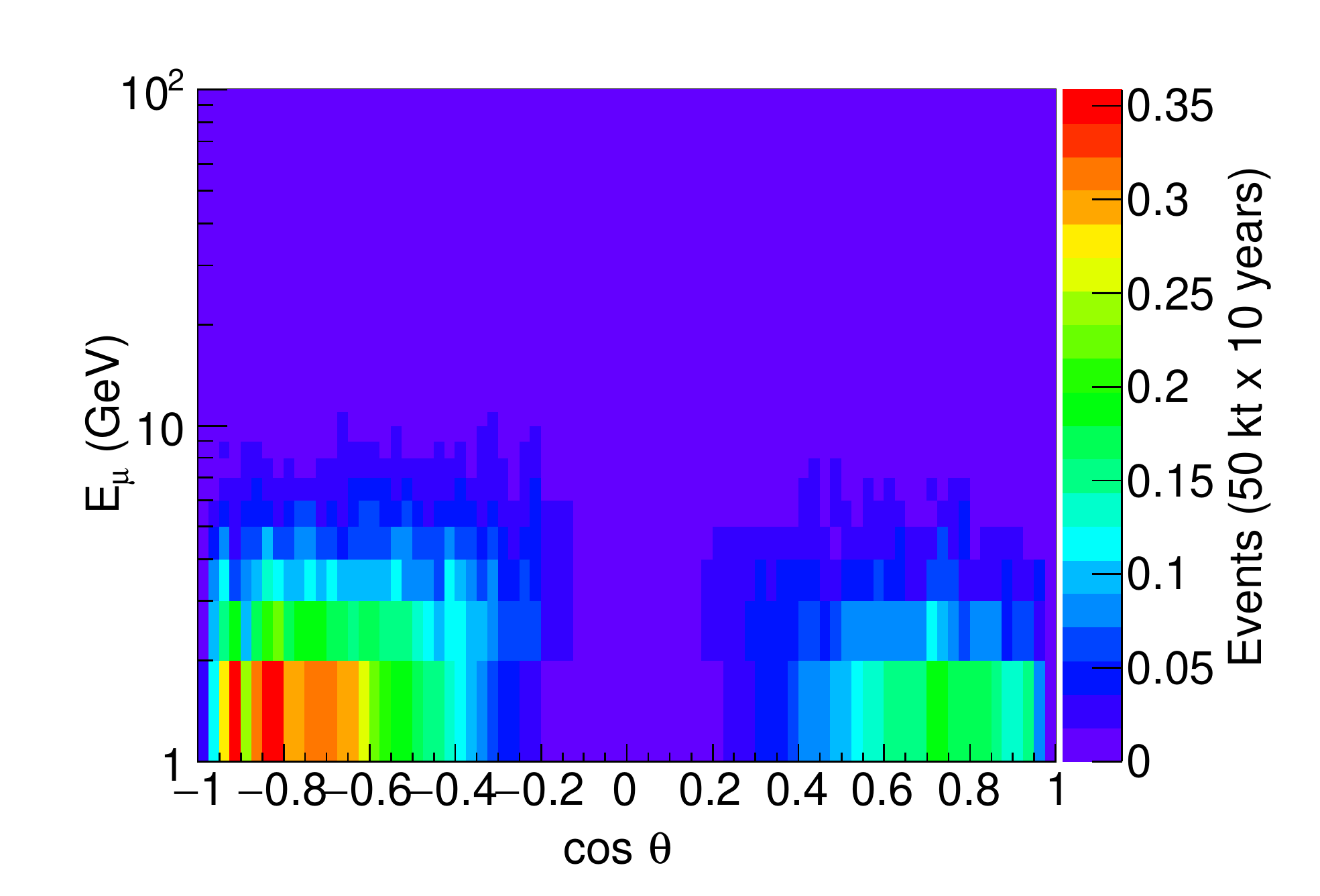}
\caption{\label{fig:7} The $\mu^{-}$ event distribution at ICAL due to atmospheric background after applying the 
angular suppression corresponding to a 100 GeV WIMP. $\cos\theta = 1$ represents upward going muons. The plot is for 10 years of ICAL running.}
\end{figure}


\begin{figure}[tbp]
     \subfigure[]{\includegraphics[width=0.48\textwidth]{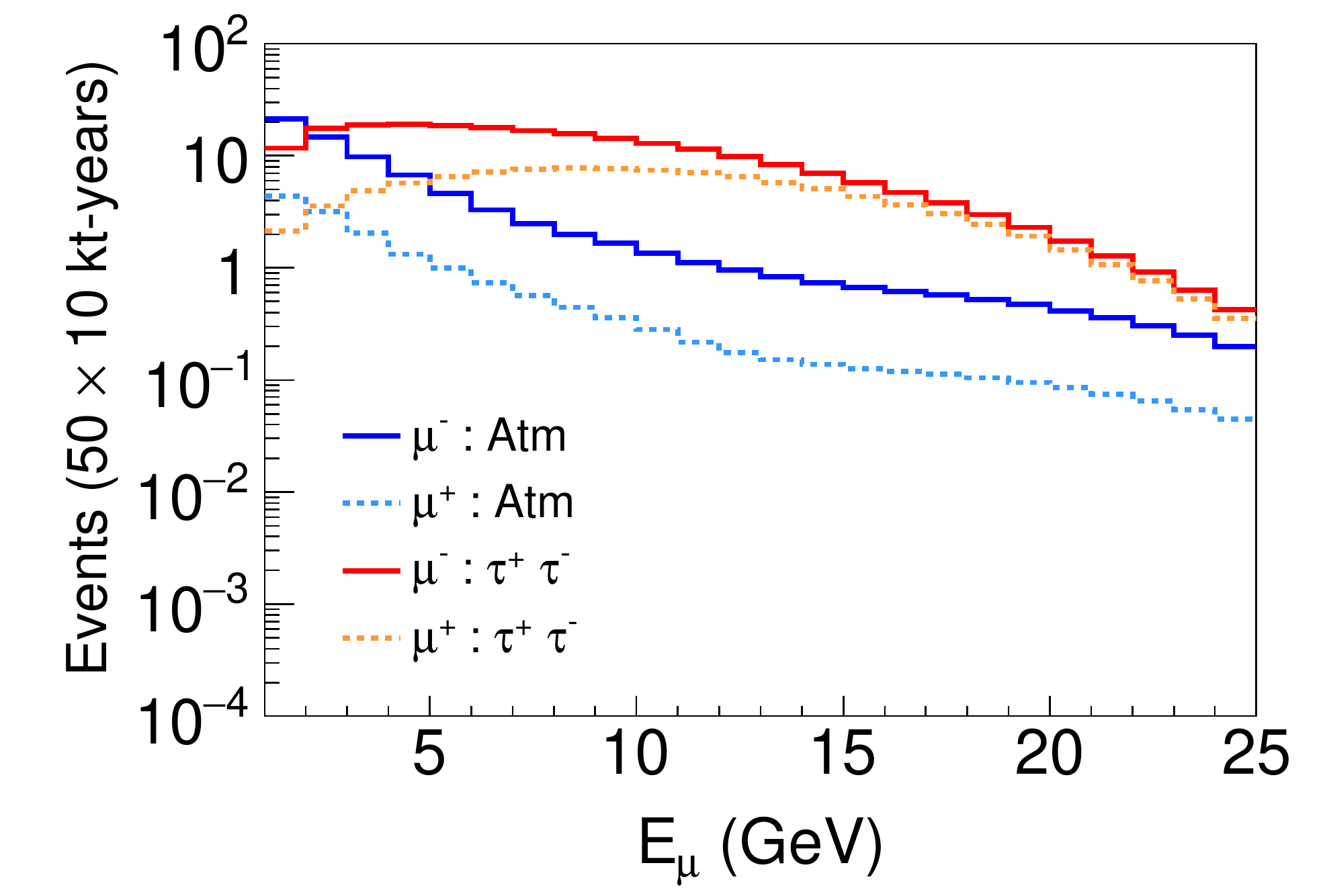}\label{fig:a}}
     \subfigure[]{\includegraphics[width=0.48\textwidth]{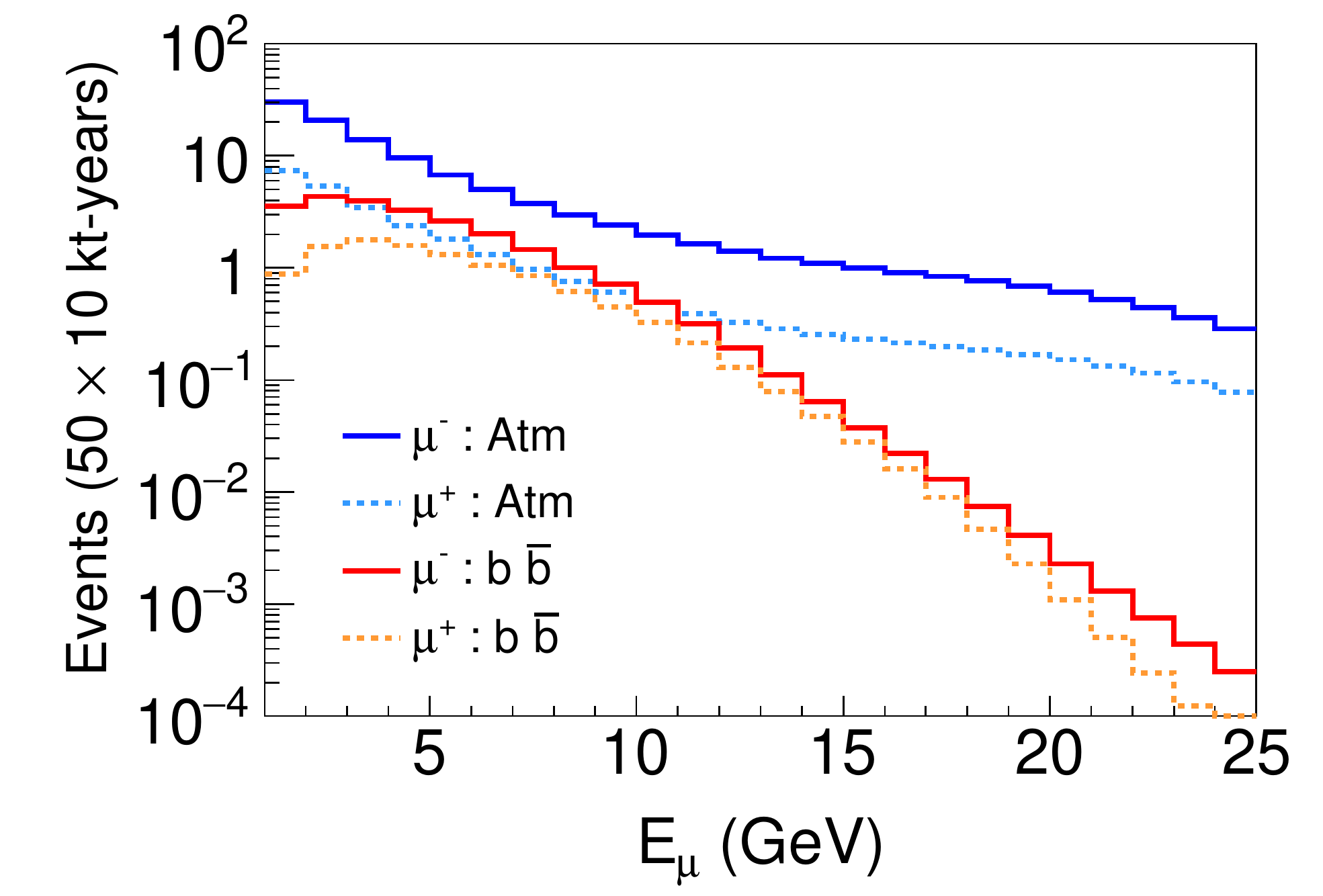}\label{fig:b}}
     \subfigure[]{\includegraphics[width=0.48\textwidth]{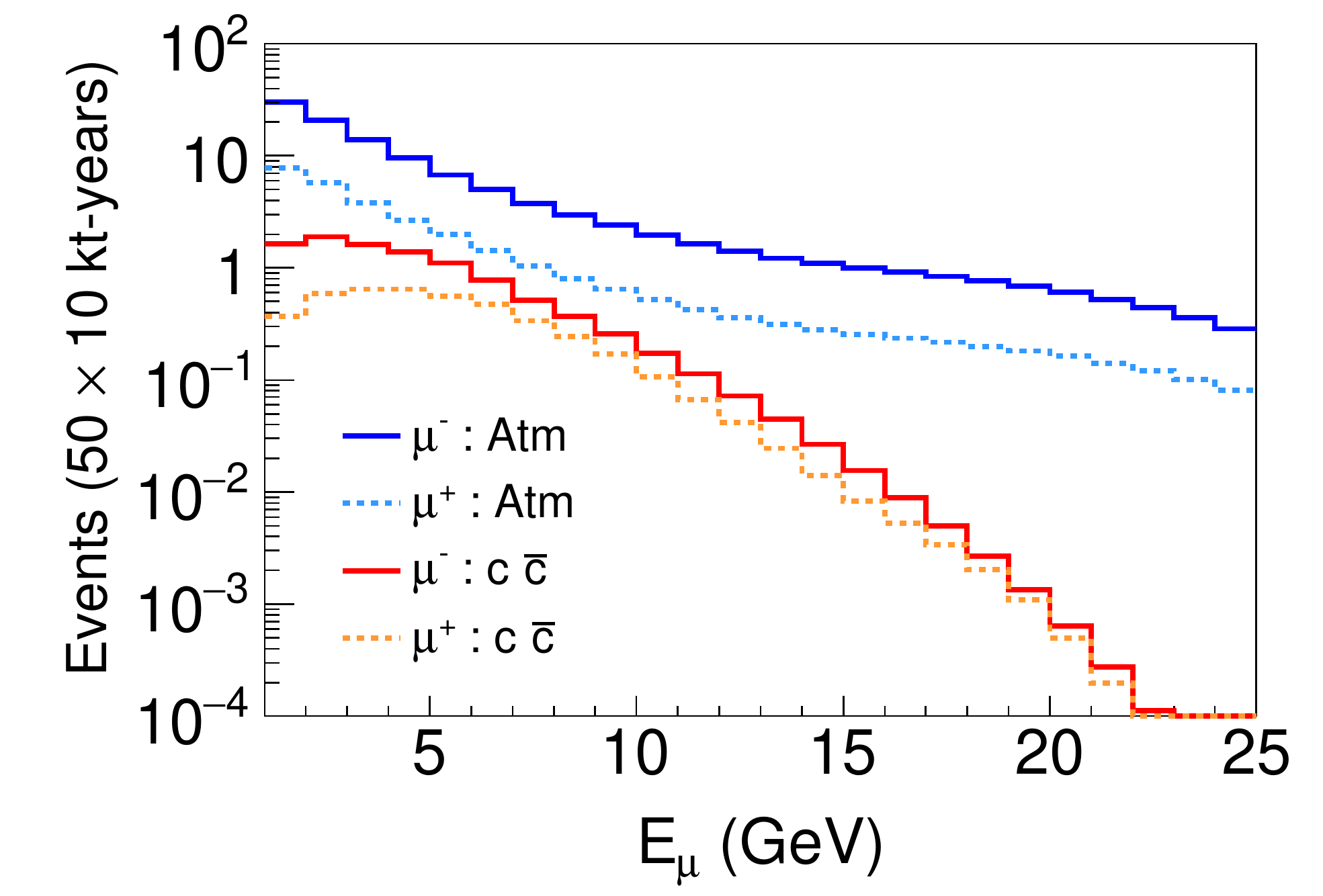}\label{fig:c}}
     \subfigure[]{\includegraphics[width=0.48\textwidth]{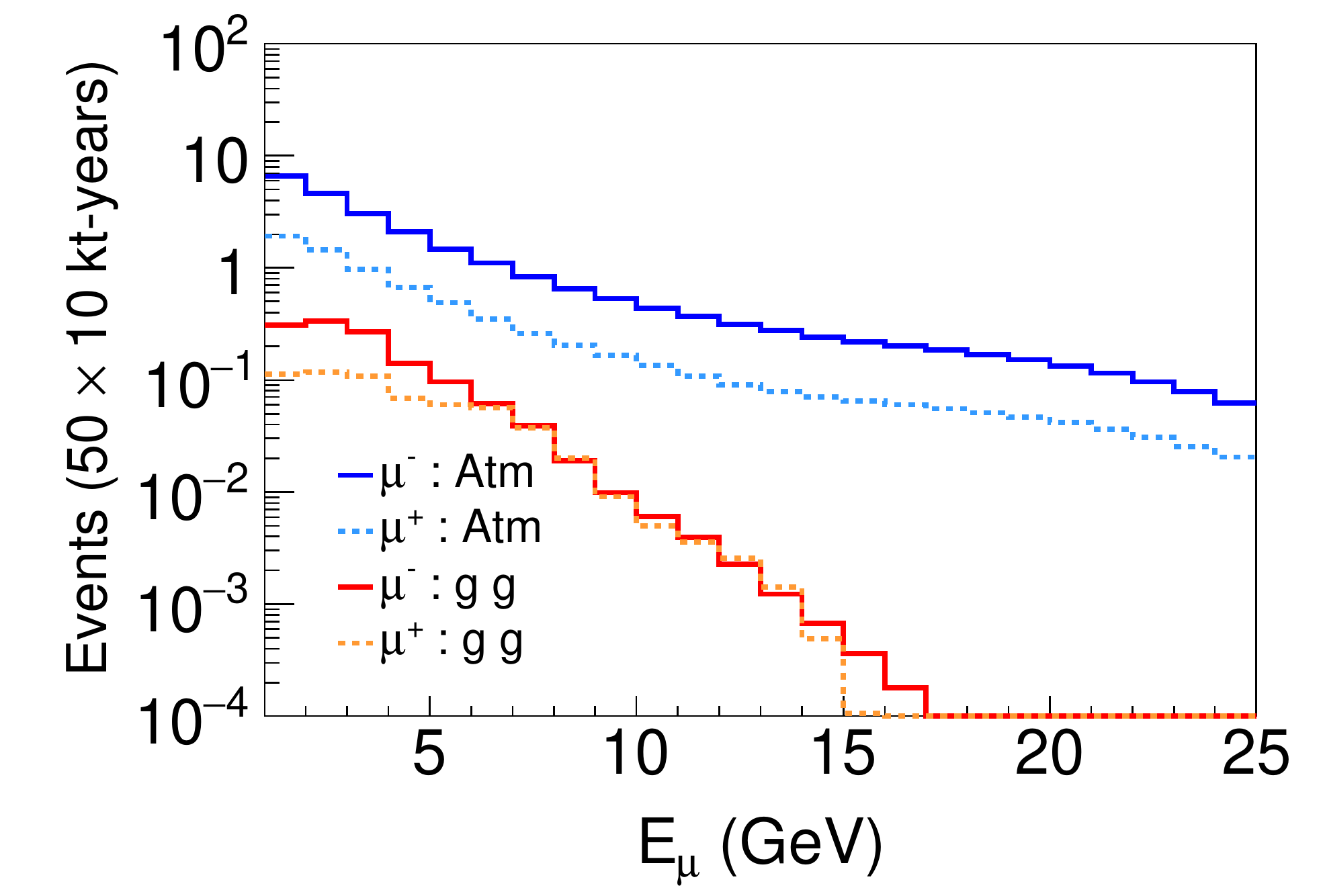}\label{fig:d}}
     \subfigure[]{\includegraphics[width=0.48\textwidth]{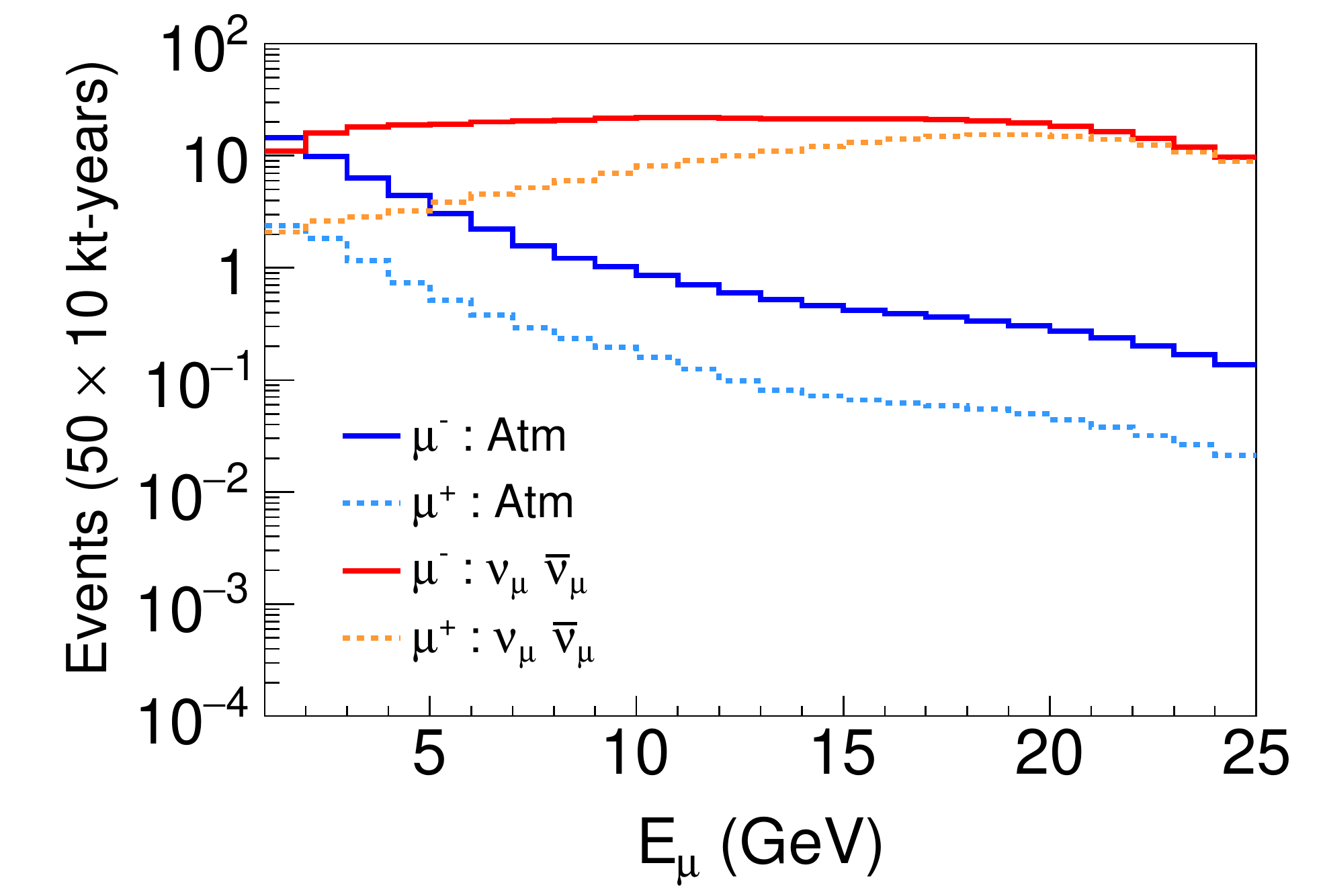}\label{fig:e}}
     \subfigure[]{\includegraphics[width=0.48\textwidth]{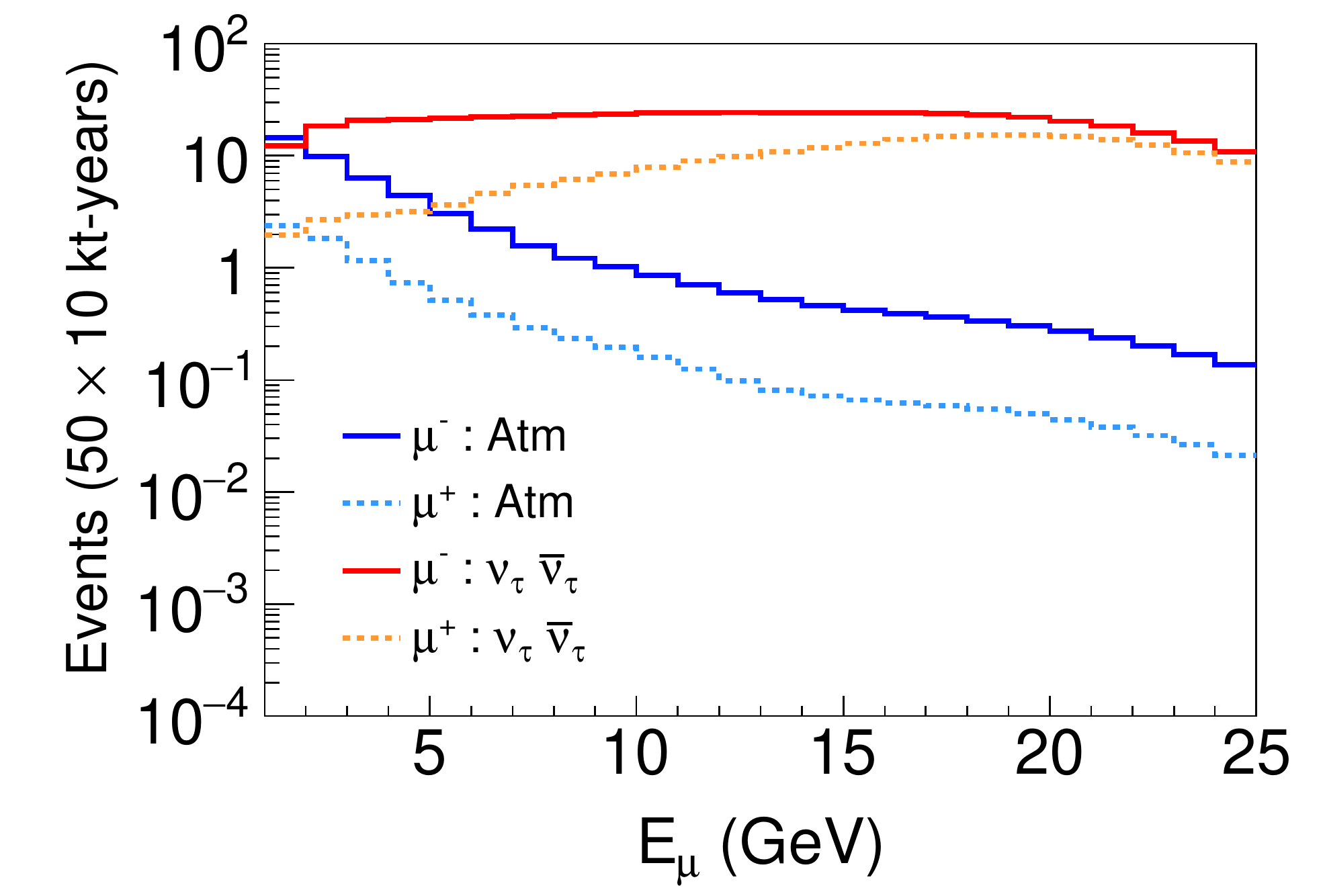}\label{fig:f}}
\caption{\label{fig:9} The $\mu^-$ (red solid lines) and $\mu^+$ (orange dotted lines) event distribution at ICAL due to signal neutrinos arising out of WIMP annihilations in the sun through various annihilation channels . The signal events correspond to neutrino fluxes arising due to SD capture rate. A cross-section of $\sigma_{SD} =10^{-39} cm^{2}$ has been assumed for the signal neutrinos and $m_\chi$ is taken as 25 GeV. Also shown are $\mu^-$ (blue solid lines) and $\mu^+$ (azure dotted lines) event distributions due the reduced atmospheric neutrino background after applying the cone-cut angular suppression and the solar exposure function suppression. }
\end{figure}

After having applied the cone cut to each of the atmospheric neutrino background events, we have to assure that the events are indeed coming from the direction of the sun. For the case of WIMP annihilation in the sun, the direction of the signal neutrinos is same as the direction of the sun. Since we are using a Monte-Carlo generated data for our analysis, we assign a weight to each of the background events, where the weight corresponds to the probability that a particular event has originated from the direction of the sun. We call the probability of the sun exposing a particular zenith and azimuth, for a given latitude and longitude at the earth, as the solar exposure probability. We calculate this using WIMPSIM which uses SLALIB \cite{2014ascl.soft03025W} routines. Fig.~\ref{fig:6} shows the solar exposure probability for the location of INO. The neutrino fluxes arising due to WIMP annihilations in the sun necessarily follow the same angular distribution. The events accepted after this step are the final background events which are shown in  Fig.~\ref{fig:7}. The reduced background has been shown for a the cone-cut angle $\theta_{90}$ corresponding to a 100 GeV WIMP and the $\tau^{+} \tau^{-}$ channel. \\ 

To compare our WIMP indirect detection signal with the reduced atmospheric neutrino background in ICAL, we show in Fig.~\ref{fig:9} the $\mu^-$ and $\mu^+$ signal events due to a 25 GeV WIMP annihilating into $\tau^{+} \tau^{-}$, $b \bar{b}$, $c \bar{c}$, $b \bar{b}$,$\nu_{\mu}\bar\nu_{\mu},\nu_{\tau} \bar\nu_{\tau}$ and $g g$ channels. $\nu_{e}\bar\nu_{e}$ event spectra is similar to  $\nu_{\mu}\bar \nu_{\mu}$ and $\nu_{\tau} \bar\nu_{\tau}$ and hence we have not shown it here. Also shown are corresponding reduced atmospheric neutrino background events. In Fig.~\ref{fig:a}, we show the signal events for the $\tau^{+} \tau^{-}$ channel for a 25 GeV WIMP and the atmospheric neutrino background corresponding to the neutrino and antineutrino $\theta_{90}$ for this channel and 25 GeV WIMP mass. Like-wise, Fig.~\ref{fig:b} shows the corresponding events spectra and atmospheric neutrino background expected for the $b \bar{b}$ channel. As noted before, since the neutrino flux from the $\tau^{+} \tau^{-}$ channel is higher than the $b \bar{b}$ channel and since it also produces a harder neutrino spectrum, the signal event spectrum is higher for the former as well as harder for both neutrinos as well as antineutrinos. The atmospheric neutrino background too is lower for the $\tau^{+} \tau^{-}$ channel since a harder neutrino spectrum gives a smaller $\theta_{90}$, improving the cone-cut background rejection employed in our analysis. Therefore, the signal to background ratio and hence the sensitivity to WIMP is expected to be better for the $\tau^{+} \tau^{-}$ channel, as we will see in the following sections. Fig.~\ref{fig:c} shows the events due to $c \bar{c}$ annihilation channel. Since the fluxes from $c \bar{c}$ are only slightly lower in comparision to $b \bar{b}$, the event distribution follows the similar trend. However, since the $\theta_{90}$ values for $c \bar{c}$ and $b \bar{b}$ channels are almost comparable for various WIMP masses, it would be difficult to distinguish between these two channels. The events due to direct annihilation of WIMP into neutrinos give rise to a higher number of event and hence better sensitivity to indirect detection.



\section{The statistical analysis}
\label{sec:analysis}

We perform a $\chi^{2}$ analysis to obtain expected sensitivity limits on SD and SI WIMP-nucleon scattering cross-sections for given WIMP masses. We simulate the prospective data at ICAL for no WIMP annihilation and fit it with a theory where WIMP annihilate in the sun to give neutrinos. Therefore, the ``data" or ``observed" events correspond to only the reduced atmospheric neutrino backgrounds, whereas the ``theory" or ``predicted" events comprise the sum of the signal events due to WIMP annihilation in the sun as well as the atmospheric neutrino background events. The $\chi^2$ function is defined as: 
\begin{equation}
 \chi^2 = \chi^2(\mu^-) +  \chi^2(\mu^+)
 \,,
\end{equation}
where 
\begin{equation}
\chi^2(\mu^\pm) = \min_{\xi^\pm_k}\sum_{i=1}^{N_i}\sum_{j=1}^{N_j}
\bigg [ 2\bigg(N_{ij}^{\rm th}(\mu^\pm) - N_{ij}^{\rm ex}(\mu^\pm)\bigg ) + 
2N_{ij}^{\rm ex}(\mu^\pm)\ln\bigg(\frac{N_{ij}^{\rm ex}(\mu^\pm)}{N_{ij}^{\rm th}(\mu^\pm)}
\bigg ) \bigg]
+ \sum_{k=1}^l {\xi^\pm_k}^2
\,,
\end{equation}
\begin{equation}
N_{ij}^{\rm th}(\mu^\pm) = {N'}_{ij}^{\rm th}(\mu^\pm)\bigg(1+\sum_{k=1}^l \pi_{ij}^k{\xi^\pm_k}\bigg)  
+{\cal O}({\xi^\pm_k}^2)\,,
\end{equation}
${N'}_{ij}^{\rm th}(\mu^\pm)$ and $N_{ij}^{\rm ex}(\mu^\pm)$ are the `predicted' and `observed' number of $\mu^\pm$ events at ICAL respectively. As explained above, in our analysis $N_{ij}^{\rm ex}(\mu^\pm)$ include only the reduced atmospheric neutrino background events, while ${N'}_{ij}^{\rm th}(\mu^\pm)$ include both signal events from WIMP annihilation in the sun as well as the reduced background events from atmospheric neutrinos. The quantities $\pi_{ij}^k$ are the correction factors due to the $k^{th}$ systematic uncertainty, and $\xi^\pm_k$ are the corresponding pull parameters. Since this is an analysis which looks for signal events above atmospheric neutrino backgrounds, and it is well known that there are substantial systematic uncertainties in the predicted atmospheric neutrino fluxes, we include systematic uncertainties on the atmospheric neutrino background. Like our previous analysis \cite{Choubey:2015xha}, we include 5 systematic errors on the atmospheric neutrino background: 20~\% error on flux normalisation, 10~\% error on cross-section, 5~\% uncorrelated error on the zenith angle distribution of atmospheric neutrino fluxes and 5~\% tilt error. We further include a  5~\% overall error to account for detector systematics\footnote{Simulations to estimate the detector systematic uncertainties in ICAL is underway. This number could therefore change when better estimates of this become available.}. The individual contributions from $\mu^-$ and $\mu^+$ data samples are calculated by minimising over the pull parameters. These are then added to obtain the $\chi^2$ for a given set of WIMP mass and WIMP-nucleon cross-section.
\section{Results}
\label{sec:results} 

\begin{figure}[tbp]
\centering 
\includegraphics[width=.8\textwidth]{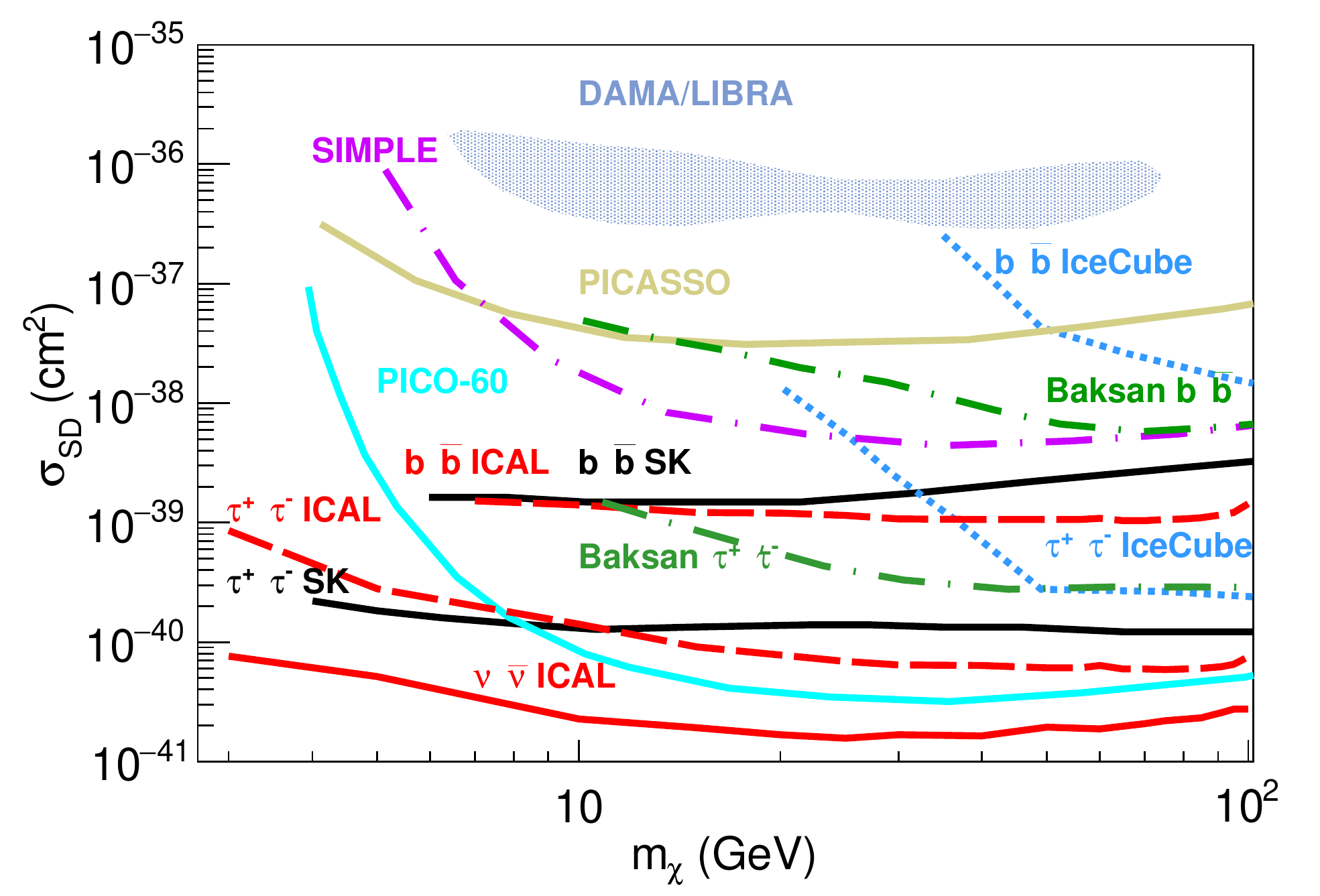}
\caption{\label{fig:13}  The expected 90 \%  C.L. sensitivity limit on the WIMP-nucleon spin-dependent cross-sections as a function of the WIMP mass. The ICAL expected sensitivity are shown for $\nu \bar{\nu}$ (red solid line), $\tau^{+} \tau^{-}$ (red dashed line) and $b \bar{b}$ (red dashed line) channels and for 10 years of running of ICAL. Current 90 \% C.L. limits from other indirect detection and direct detection experiments have been shown. Also shown is the region compatible with the claimed signal seen by DAMA/LIBRA. }
\end{figure}

\begin{figure}[tbp]
\centering
\includegraphics[width=.8\textwidth,origin=c]{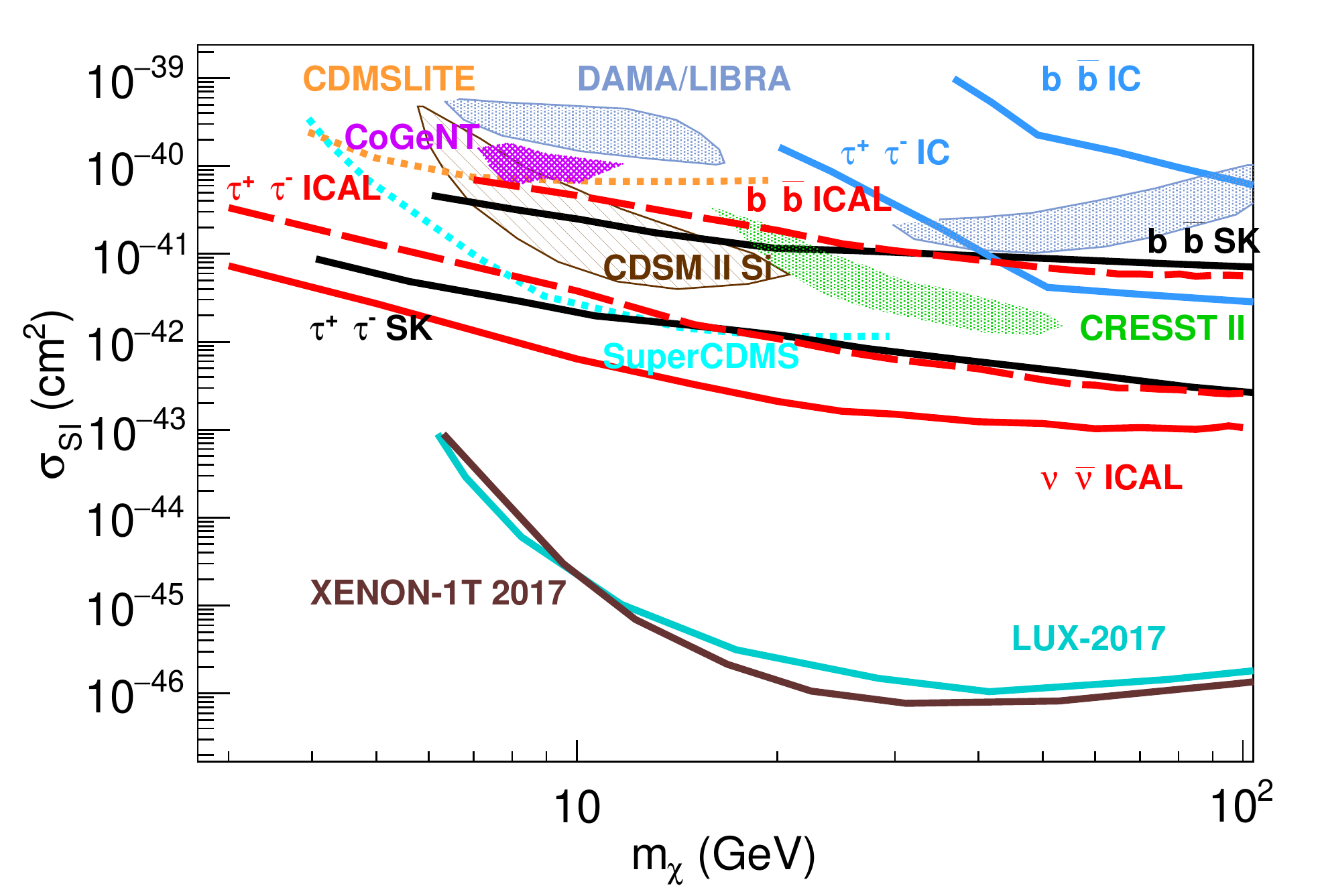}
\caption{\label{fig:14} The expected 90 \%  C.L. sensitivity limit on the WIMP-nucleon spin-independent cross-sections as a function of WIMP mass. The expected ICAL sensitivity are shown for $\nu \bar{\nu}$ (red solid line), $\tau^{+} \tau^{-}$ (red dashed line) and $b \bar{b}$ (red dashed line) channels and for 10 years of running of ICAL. Current 90 \% C.L. limits from other indirect detection and direct detection experiments have been shown. Also shown is the region compatible with the claimed signal seen by DAMA/LIBRA, CoGeNT, CRESSTII and CDMS II Si. }
\end{figure}

\begin{figure}[tbp]
\centering 
\includegraphics[width=.8\textwidth]{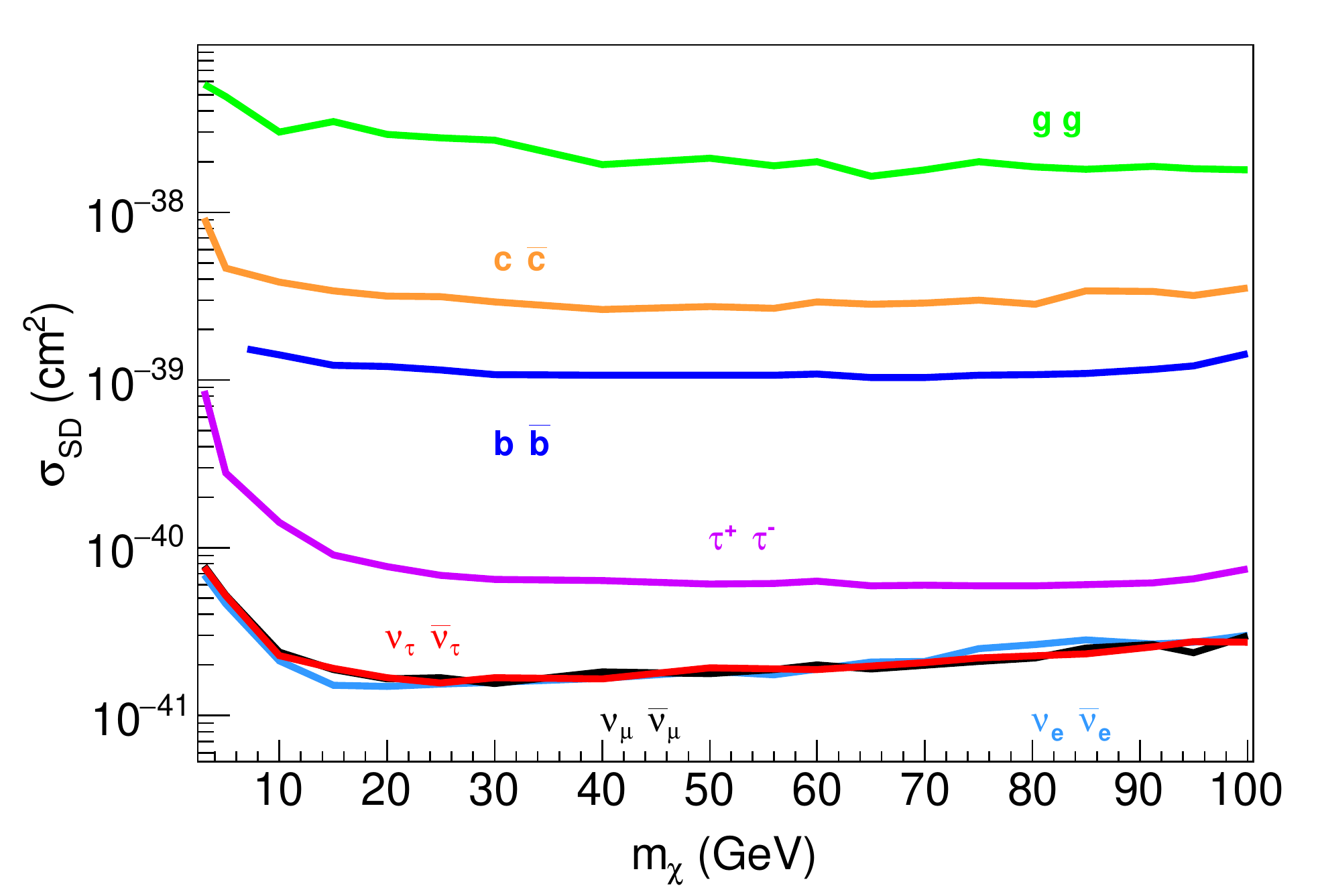}
\caption{\label{fig:12a} The expected 90 \%  C.L. sensitivity limit for ICAL on the WIMP-nucleon spin-dependent cross-sections as a function of the WIMP mass and for different annihilation channels.}
\end{figure}

\begin{figure}[tbp]
\centering 
\includegraphics[width=.8\textwidth]{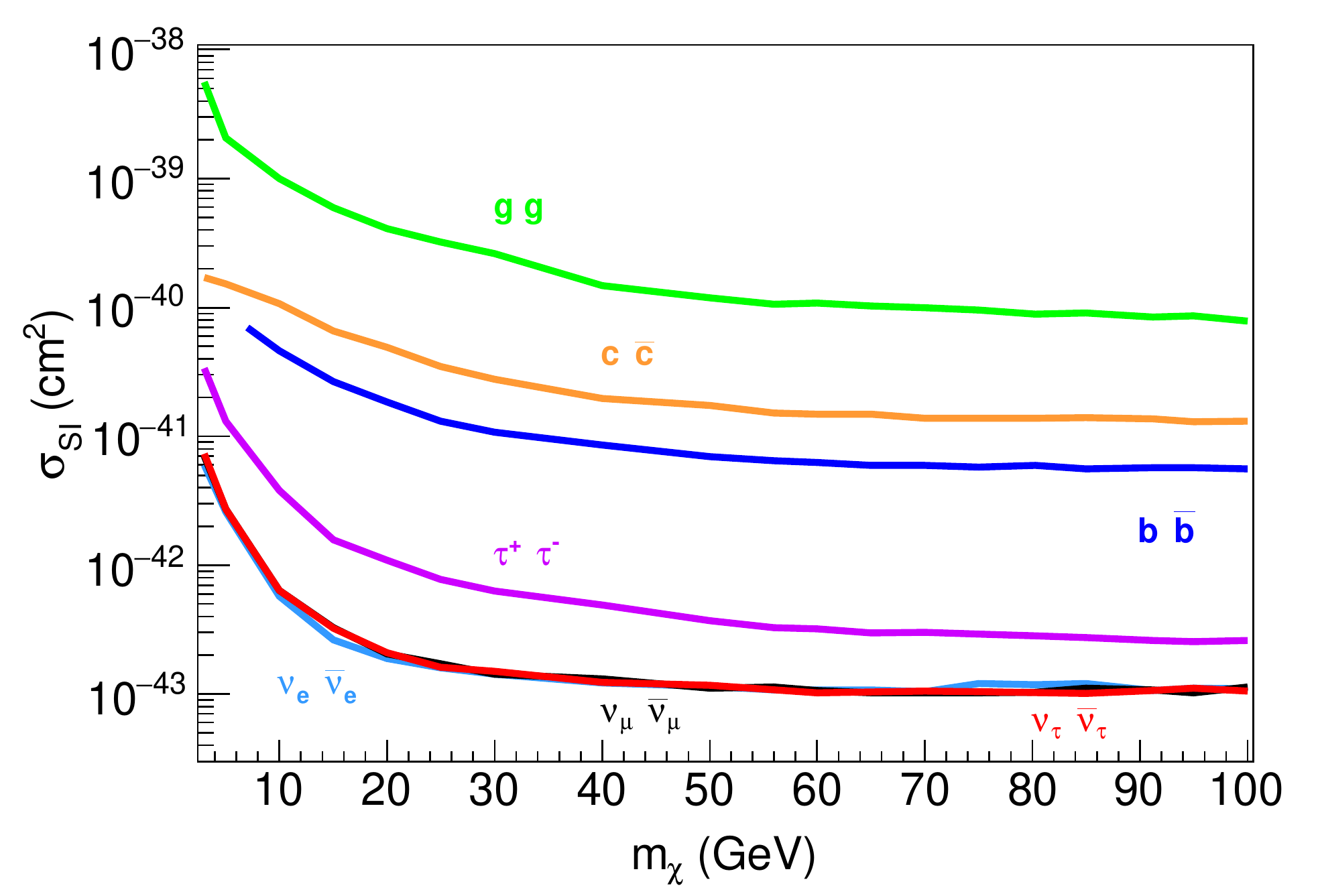}
\caption{\label{fig:12b} The expected 90 \%  C.L. sensitivity limit for ICAL on the WIMP-nucleon spin-independent cross-sections as a function of the WIMP mass and for different annihilation channels.}
\end{figure}

\begin{figure}[tbp]
\centering 
\includegraphics[width=.8\textwidth]{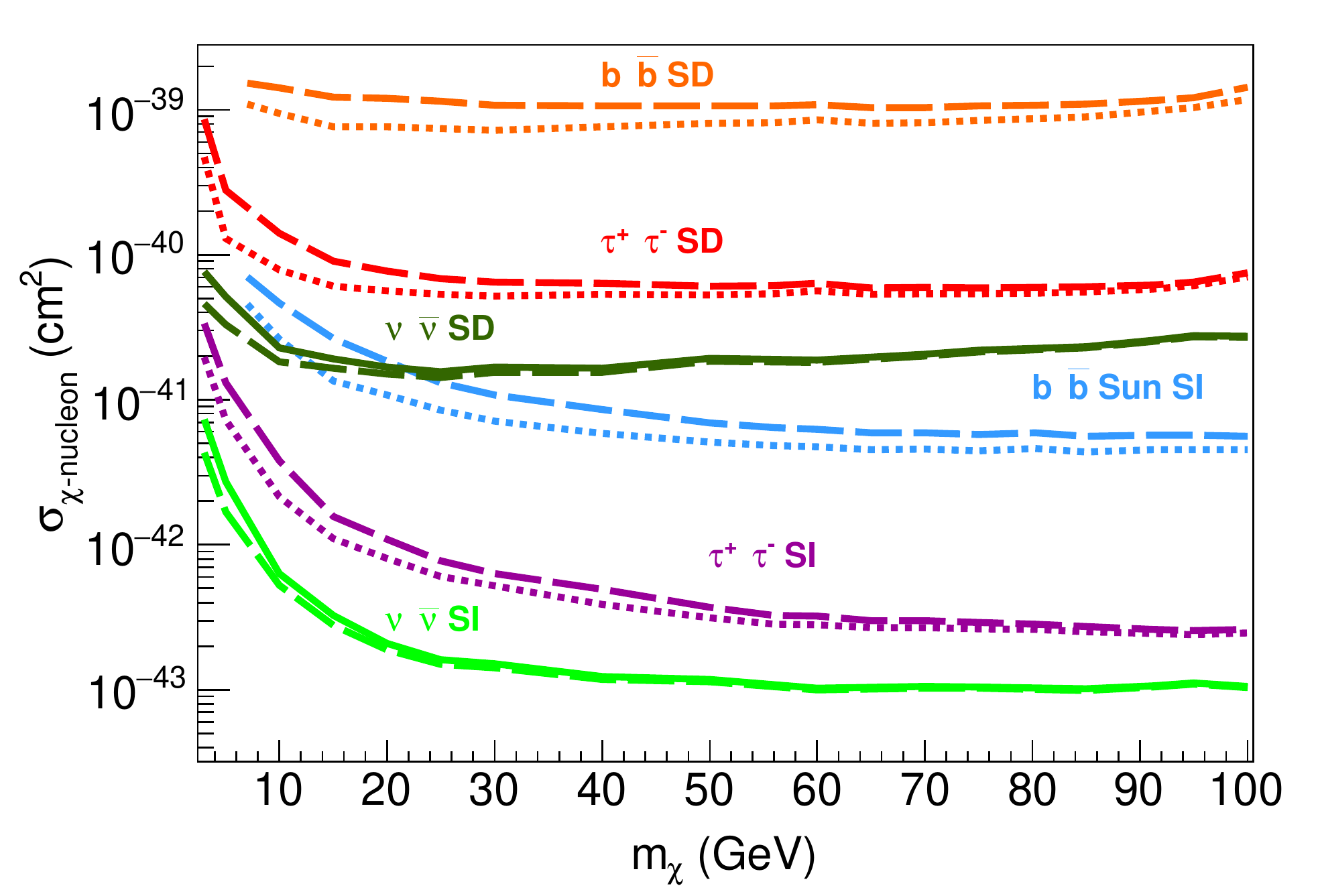}
\caption{\label{fig:12c} The expected 90 \%  C.L. sensitivity limit for ICAL on the WIMP-nucleon spin-dependent and spin-independent cross-sections as a function of the WIMP mass and for different annihilation channels. The solid lines are the sensitivity limits calculated using detector systematics as described in Sec~\ref{sec:analysis}. The corresponding dotted lines are without systematics.}
\end{figure}

\begin{figure}[tbp]
\centering 
\includegraphics[width=.8\textwidth]{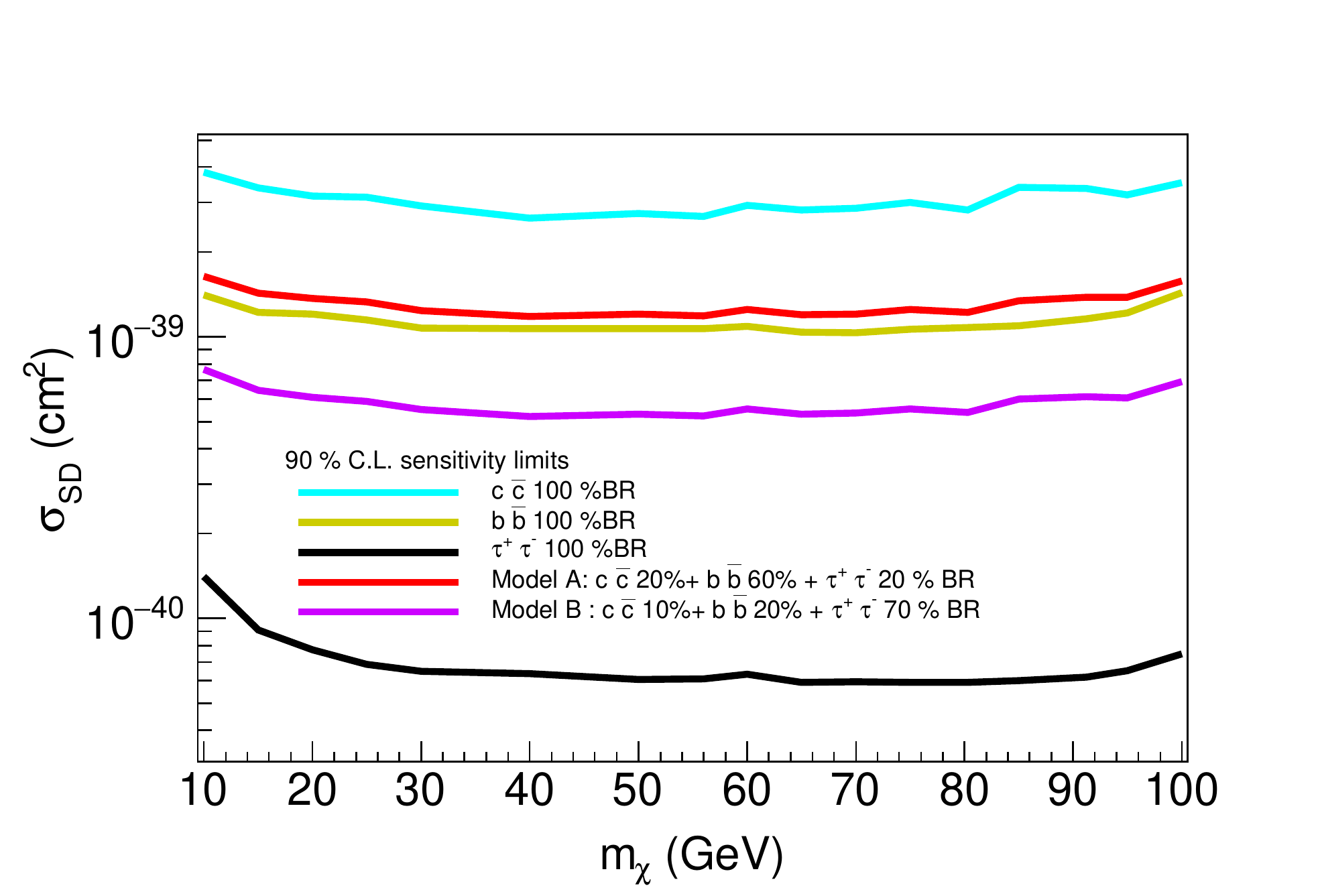}
\caption{\label{fig:12d} Impact of varying branching ratios on the expected 90 \%  C.L. sensitivity limit at ICAL.  The shown limits are for WIMP-nucleon spin-dependent cross-sections as a function of the WIMP mass and for three annihilation channels (same as in Fig.~\ref{fig:13}). Detector systematics have been included. The expected 90 \%  C.L. sensitivity limit due to toy Models A (red line) and B (violet line) have been shown.}
\end{figure}

We now present our main results on the expected sensitivity of ICAL to indirect detection of dark matter. The 90 \% C.L. expected  sensitivity from 10 years of running of ICAL is shown in Fig.~\ref{fig:13} in the $\sigma_{SD}-m_\chi$ plane for the spin dependent cross-section. The expected sensitivity assuming 100 \% BR in $\nu \bar{\nu}$ (red solid line), $\tau^+\tau^-$ (red dashed line) and $b\bar b$ (red dashed line) channels have been shown. Also shown are limits from other complementary indirect detection experiments: IceCube \cite{Aartsen:2012kia} $\tau^{+}\tau^{-}$ (bottom, light blue and dotted) and $b \bar{b}$ (top, light blue and dotted) channels, Super-Kamiokande (SK) \cite{Choi:2015ara} $\tau^{+}\tau^{-}$ (bottom, black) and $b \bar{b}$ (top, black), BAKSAN \cite{Boliev:2013ai} $\tau^{+}\tau^{-}$ (bottom, green dot dashed) and $b \bar{b}$ (top, green dot dashed) channels. For comparison we also show the 90 \% C.L. limits from direct detection experiments: PICASSO \cite{Archambault:2012pm} (brownish-yellow), SIMPLE \cite{Felizardo:2011uw} (violet long dot dashed) and PICO-60 $\textup{C}_{3}\textup{F}_{8}$ \cite{Amole:2017dex} (cyan solid line). Finally, the blue-gray shaded region is the $3\sigma$ C.L. area compatible with the signal claimed by DAMA/LIBRA \cite{Savage:2008er,Bernabei:2008yi}. We find that the expected sensitivity from 10 years of running of ICAL is comparable\footnote{We compared the event spectra for $\tau^{+} \tau^{-}$ and atmospheric background as given in Fig.~\ref{fig:a} with the corresponding signal and background event spectra of Fig. 2 of \cite{Agarwalla:2011yy}. Applying the same statistical analysis on both these event sets, the calculated sensitivity limits were found to be comparable.} to SK for both $\tau^+\tau^-$ and $b\bar b$ channels, with ICAL performing a tad better for all WIMP masses greater than 10 GeV. Note that ICAL is expected to be better than all other direct as well as indirect experiments which have placed limits on the WIMP-nucleon spin-dependent scattering cross-section. In particular, the sensitivity of the ICAL experiment is expected to be better than the current limits from IceCube and Baksan. The limits on $\sigma_{SD}$ from the direct detection experiments, in general, are weaker than those from indirect detection experiments for all ranges of WIMP masses $m_\chi$ with the exception of the limit obtained by PICO-60 $\textup{C}_{3}\textup{F}_{8}$.\\

The 90 \% C.L. expected  sensitivity from 10 years of running of ICAL for the spin-independent cross-section is shown in the $\sigma_{SI}-m_\chi$ plane in Fig.~\ref{fig:14} along with limits from other experiments. We show the expected sensitivity assuming 100 \% BR in $\nu_{\tau} \bar{\nu}_{\tau}$ (red solid line), $\tau^+\tau^-$ (red dashed line) and $b\bar b$ (red dashed line) channels. Also  shown are the current limits from earlier and on-going experiments. For the SI cross-sections, the limits from the direct detection experiments are significantly better than those from the indirect detection experiments. Allowed regions from the direct searches claiming positive signal for dark matter at DAMA/LIBRA \cite{Savage:2008er,Bernabei:2008yi} is shown by metallic blue shaded region at $3\sigma$ C.L., at CoGeNT \cite{Aalseth:2010vx} by violet diagonally cross-hatched region at 90 \% C.L., at CRESSTII \cite{Angloher:2011uu} by the green shaded region at $2\sigma$ C.L. and  at CDMS II Si \cite{Agnese:2013rvf} by brown hatched region at 90 \% C.L. The limits on SI cross-sections from direct detection experiments are shown for SuperCDMS \cite{Agnese:2014aze} (cyan dotted line), CDMSlite \cite{Agnese:2013jaa} (orange dotted), XENON-1T \cite{Aprile:2017iyp} (brown solid line) and LUX \cite{Akerib:2016vxi} (solid cyan line). The limits from Xenon-1T are currently the best limits on WIMP-nucleon SI cross-sections. Also shown are the less competitive limits from the indirect searches at IceCube \cite{Aartsen:2012kia} for $\tau^{+}\tau^{-}$ (cyan solid line) and $b \bar{b}$ (cyan dotted line) channels, and SK \cite{Choi:2015ara} for $\tau^{+}\tau^{-}$ (dark violet line) and $b \bar{b}$ (light violet line) channels. As for the SD cross-section case, we find that the expected sensitivity of ICAL is comparable to that from SK for both channels and better than the current limits from IceCube. It is worth pointing out that even though the limits on SI case from indirect detection experiments are expected to be poorer than from direct detection experiments, they provide an independent check on the WIMP parameters and can be used as a complementary probe of the WIMP paradigm. \\

Now we show the expected sensitivity of ICAL to indirect detection due to several WIMP annihilation channels. The 90 \% C.L. expected  sensitivity from 10 years of running of ICAL is shown in Fig.~\ref{fig:12a} in the $\sigma_{SD}-m_\chi$ plane for the spin dependent cross-section. The green lines are for $g g$ channel, orange lines for $c \bar{c}$, blue for $b\bar b$, violet for $\tau^+\tau^-$, azure lines for $\nu_{e} \bar{\nu}_{e}$, black for  $\nu_{\mu} \bar{\nu}_{\mu}$ and red for $\nu_{\tau} \bar{\nu}_{\tau}$ channel. Fig.~\ref{fig:12b} presents the 90 \% C.L. expected  sensitivity from 10 years of running of ICAL for the spin-independent cross-section, shown in the $\sigma_{SI}-m_\chi$ plane for different annihilation channels. The colour coding remains the same. For the reasons described in \ref{sec:capture}, the expected sensitivity limit due to WIMP annihilating into neutrino-antineutrino channels is the strongest and weakest for the $g g$ channel. This holds for both SD and SI case.\\

In Fig.~\ref{fig:12c} we present the impact of systematic uncertainties in the atmospheric neutrino background on the indirect detection sensitivity of ICAL. The solid lines in this figure show the 90 \% C.L. expected sensitivity of ICAL when systematic uncertainties are included. These were the expected limits shown in Figs. \ref{fig:13}, \ref{fig:14}, \ref{fig:12a} and \ref{fig:12b}. The corresponding dashed lines are obtained by switching off the systematic uncertainties in the analysis. The orange lines are for $\sigma_{SD}$ and $b\bar b$ channel, the red lines are for $\sigma_{SD}$ and $\tau^+\tau^-$ channel, the azure lines are for $\sigma_{SI}$ and the $b\bar b$ channel, the magenta lines are for $\sigma_{SI}$ and $\tau^+\tau^-$ channel, the dark green lines for $\sigma_{SD}$ and $\nu \bar \nu$ channel and light green lines are for $\sigma_{SI}$ and $\nu \bar \nu$ channel. The impact of the systematic uncertainties for all the channels are seen to be more for lower WIMP masses. This is because the atmospheric neutrinos peak at lower energies and their fluxes fall as roughly $E_\nu^{-2.7}$. Therefore, for lower WIMP masses, since the neutrino spectrum from WIMP annihilation are softer, these get more affected by the uncertainties in the atmospheric neutrino fluxes. The impact of the uncertainties is also seen to be more for the $b\bar b$ channel. The reason for this behaviour is again the same. We had seen in section \ref{sec:capture} that the neutrino spectrum from the $b\bar b$ channel is softer. Similarly, the harder channels like $\nu \bar{\nu}$ and $\tau^+\tau^-$ have greater high energetic neutrino content. Therefore, the impact of the atmospheric neutrino background and also the uncertainty on the atmospheric neutrino background affect the softer channels more than the harder channels. \\

Finally, in Fig.~\ref{fig:12d} we illustrate the impact of varying branching ratios on sensitivity limits. It is clear from Fig.~\ref{fig:2} that some channels are weaker in comparision to others and would yield a weaker sensitivity to indirect detection as shown in Fig.~\ref{fig:12a} and Fig.~\ref{fig:12b}. However, these sensitivity limits are assuming $100\%$ branching ratios for each of the annihilation channel. Therefore, the above sensitivity limit indicates the best limit one would expect at ICAL due to a particular channel considered. In nature, however, we would have a mixture of fluxes from different channels and correspondingly the sensitivity limits would be altered. For a given WIMP model, the branching ratio of a particular channel would be known and corresponding contribution in terms of neutrino fluxes can be easily calculated. In a framework of a particular model, therefore, the sensitivity due to this mixture of fluxes would get scaled appropriately. To illustrate this point, we consider two toy models and calculate their expected sensitivities at ICAL. We take two toy models restricted to three annihilation channels: Model A which has a following prediction : $BR_{c \bar c}: BR_{b \bar b}: BR_{\tau^{+} \tau^{-}} = 20\%:60\%:20\%$ and Model B which predicts $BR_{c \bar c}: BR_{b \bar b}: BR_{\tau^{+}\tau^{-}} = 10\%:20\%:70\%$. We see from the Fig.~\ref{fig:12d} that sensitivity due to Model B (violet line) is more than Model A (red line), and it is expected since it has greater contribution from a `harder’ channel which is $\tau^{+}\tau^{-}$ in this case. For both the models, sensitivites are indeed bounded by the strongest and the weakest channels which in this partcular case are $\tau^{+}\tau^{-}$ and $c\bar c$ respectively.

\section{Conclusions}
\label{sec:summary}

If the WIMP paradigm as a solution to the observed dark matter abundance of the universe is indeed true, they would be gravitationally captured by the sun. These WIMP would eventually accumulate in the centre of the sun, where they would annihilate into standard model particle-antiparticle pairs. All charged particles as well as photons coming from the showering of these particles would be captured in the sun, and only the neutrinos would manage to escape and reach the earth. Dark matter indirect detection experiments aim to observe these neutrinos. In this work, we probe the potential of the ICAL detector to detect the neutrinos from WIMP annihilations in the sun. This work is a part of ongoing studies to probe the physics potential of the ICAL detector. \\

We performed a study of $\mu^-$ and $\mu^+$ events arising at ICAL due to such neutrinos through various WIMP annihilation channels : $\tau^{+} \tau^{-}$, $b \bar{b}$, $c \bar{c}$, $b \bar{b}$, $\nu_{e}\bar\nu_{e},\nu_{\mu}\bar\nu_{\mu},\nu_{\tau} \bar\nu_{\tau}$ and $g g$ channels. We simulated the expected event spectrum for the dark matter signal using the event generator GENIE, suitably modified to include the ICAL detector geometry. The GENIE output is given in terms of true energy and true zenith angle of the muon and is for a 100 \% efficient ideal detector. We first fold these events with detector efficiency and charge identification efficiency. Next, in order to simulate the events in bins of measured muon energy and muon zenith angle, we fold them with the muon energy resolution functions and muon zenith angle resolution functions. We performed the ICAL detector simulation using the Geant4-based ICAL detector code to obtain the detector and charge identification efficiencies as well as the muon energy and angle resolution functions. These simulations are an extension of the detector simulations performed by the INO collaboration \cite{Chatterjee:2014vta}, where the detector response was simulated for muon energies up to 25 GeV. We extended this analysis to muon energies up to 100 GeV in order to analyse the indirect signal for higher WIMP masses. The detector and charge identification efficiencies as well as muon energy and zenith angle resolutions are functions of both true muon energy and true muon zenith angle. We presented our results on the ICAL detector response for higher energy muon in  Appendix \ref{sec:append1}. \\

The atmospheric neutrinos pose a serious background to the signal neutrinos. However, the atmospheric neutrinos come from all directions while the dark matter signal neutrinos only come from the direction of the sun. We used this feature to place an angular cut to effectively reduce the atmospheric neutrino background. We first performed a generator-level simulation to find the opening angle between the signal neutrinos and the direction of the sun such that 90 \% of the signal events were accepted. We presented these 90 \% cone-cut angles $\theta_{90}$  as a function of the WIMP mass. Heavier WIMP produce higher energy neutrinos and hence have smaller cone-cut angles. This 90 \% cone-cut criteria was then implemented on the atmospheric neutrino background, wherein the event was accepted or rejected depending on whether its zenith angle lied inside or outside the cone defined by $\theta_{90}$. Since the cone-cut angle $\theta_{90}$ was found to be different for different WIMP mass, the atmospheric neutrino events accepted  for the analysis was also different for different WIMP masses. Finally, since the sun spends a specific amount of time on a given zenith angle, we obtained the exposure function of the sun at the INO site and weighted the accepted atmospheric neutrino events with this exposure function to get the final reduced atmospheric neutrino background events as a function of the WIMP mass. We showed that the cone-cut acceptance method reduces the atmospheric neutrino background by a fact of 100. We presented the signal and background for $\tau^{+} \tau^{-}$, $b \bar{b}$, $c \bar{c}$, $b \bar{b}$, $\nu_{e}\bar\nu_{e},\nu_{\mu}\bar\nu_{\mu},\nu_{\tau} \bar\nu_{\tau}$ and $g g$ channels for WIMP mass 25 GeV and $\sigma_{SD}=10^{-39}$ cm$^2$ and showed that for the $ \tau^+ \tau^- $ and $\nu \bar{\nu}$ channels the signal is above the background for most part of the spectrum. For higher WIMP masses, the signal to background ratio were better.\\

We defined a $\chi^2$ function for the indirect detection sensitivity of ICAL to dark matter and presented the expected sensitivity in the $\sigma_{SD}-m_\chi$ and $\sigma_{SI}-m_\chi$ planes for spin-dependent and spin-independent cross-sections, respectively. The expected 90 \% C.L. sensitivity was presented for $\tau^{+} \tau^{-}$, $b \bar{b}$, $c \bar{c}$, $b \bar{b}$, $\nu_{e}\bar\nu_{e},\nu_{\mu}\bar\nu_{\mu},\nu_{\tau} \bar\nu_{\tau}$ and $g g$ channels for an exposure of 500 kt-yrs of ICAL and with systematic uncertainties on atmospheric neutrino background included in the analysis. For a WIMP mass of 25 GeV, the expected 90 \% C.L. limit using the $ \tau^+ \tau^- $ channel with 500 kt-yrs exposure in ICAL is  $\sigma_{SD} < 6.87\times 10^{-41}$ cm$^2$ and $\sigma_{SI}  < 7.75\times 10^{-43}$ cm$^2$ for the spin-dependent and spin-independent cross-sections, respectively. The effect of systematic uncertainties on the atmospheric neutrino background was also studied. \\

In conclusion, with an effective atmospheric background suppression scheme, the expected 90~\% C.L. sensitivity limits from about 10 years of running of ICAL for SD and SI WIMP-nucleon scattering cross-sections are competitive to the most stringent bounds till date. 

\acknowledgments
We gratefully acknowledge the INO collaboration for support. We sincerely thank Amol Dighe for intensive discussions. DT acknowledges K.~K.~Meghna, Moon Moon Devi, Ali Ajmi and Gobinda Majumadar for help with ICAL simulations performed in this work.
We acknowledge the HRI cluster computing facility (http://cluster.hri.res.in). The authors would like to thank the Department of Atomic Energy (DAE) Neutrino Project of Harish-Chandra Research Institute. This project has received funding from the European Union's Horizon 2020 research and innovation programme InvisiblesPlus RISE under the Marie Sklodowska-Curie grant agreement No 690575. This project has received funding from the European Union's Horizon 2020 research and innovation programme Elusives ITN under the Marie Sklodowska- Curie grant agreement No 674896. This work is also funded by Conicyt PIA/Basal FB0821.


\appendix
\section { ICAL resolution and efficiency} \label{sec:append1}

The detector efficiencies and resolution functions are needed to simulate the signal and background events in terms of reconstructed energy and zenith angle of the muon. The detector response to muons is studied using the ICAL code built on the Geant4.9.4.p02 \cite{Agostinelli:2002hh} framework. Further analyses are carried out with ROOT \cite{Brun:1997pa}. The study of the detector resolutions and efficiencies closely follows the simulation procedure carried out previously by the collaboration \cite{Chatterjee:2014vta}. However, the earlier simulation was done for muon energies of up to 20 GeV. In this work we extend the range of muon energy $E_{\mu}$ beyond 20 GeV upto 100 GeV. In our convention, the cosine of the zenith\footnote{In many places in the literature this is indeed referred to as the nadir angle, and rightly so. However, we will continue to following the convention adopted in all previous the INO simulation papers.} angle $\cos\theta =1$ represents an upward going muon, whereas $\cos\theta =-1$ indicates a downward going muon. We take 37 $E_{\mu}$ bins of variable bin-width between 1 and 100 GeV, finer bins for low energies and coarser for high energies, and 20 $\cos\theta$ bins. A large number of events are generated for a fixed  $E_{\mu}$ and $\cos\theta$,  separately for $\mu^{+}$ and $\mu^{-}$. The vertices of these events were smeared over the central region (as defined in \cite{Chatterjee:2014vta}) of ICAL. In each case, the azimuthal angle ($\phi$) was uniformly averaged over the range $-\pi\leq\phi\leq\pi$. A good reconstructed event is one that leaves a single track in the detector and crosses at least three consecutive RPC layers. A further condition $\chi^{2}/ndf <10$ is applied for choosing the reconstructed events. \\

\begin{figure}[tbp]
   \centering
     \includegraphics[width=.45\textwidth]{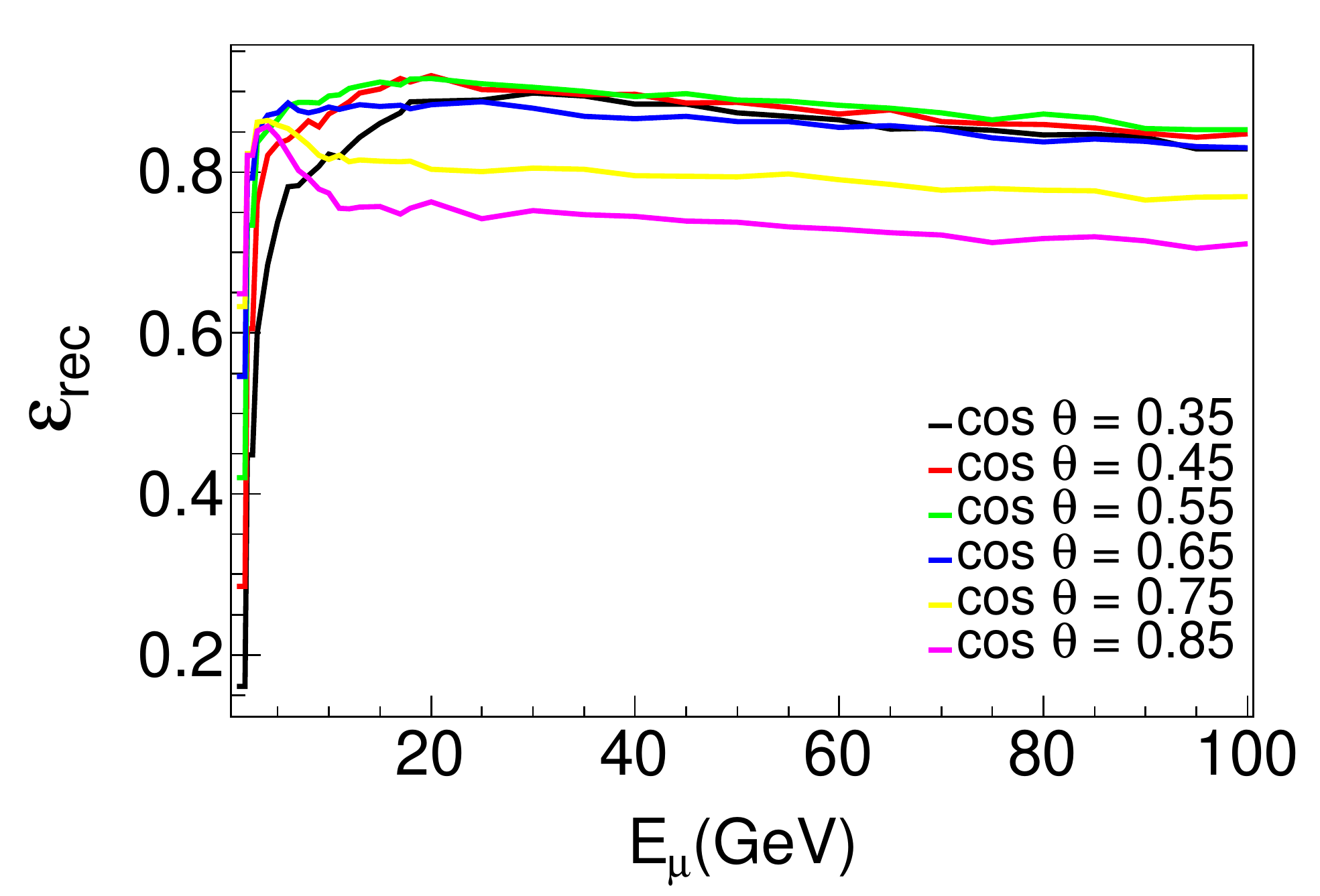}
     \hfill
     \includegraphics[width=.45\textwidth]{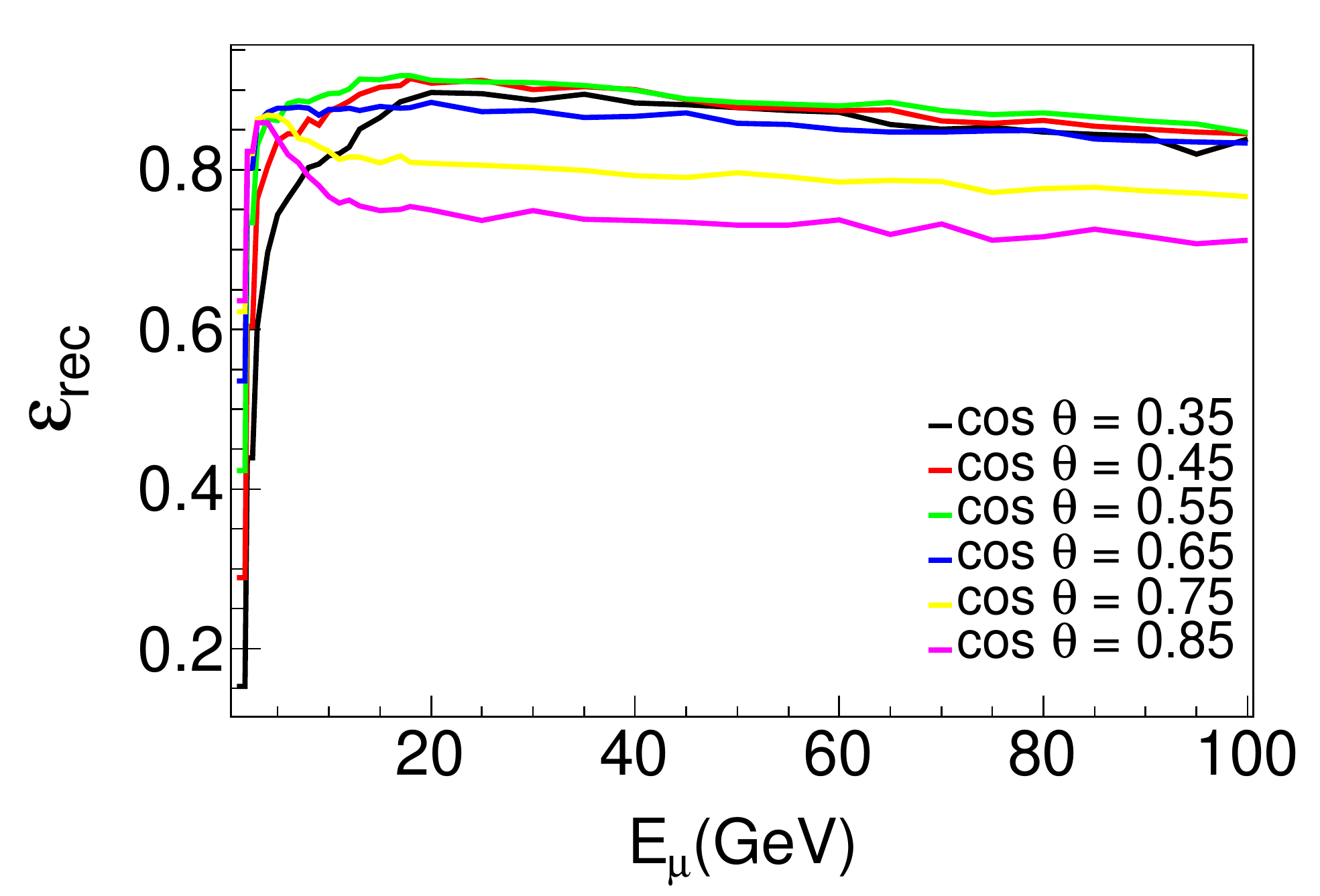}
     \caption{\label{fig:16} LEFT: Reconstruction Efficiency for $\mu^{-}$ at ICAL. RIGHT: Reconstruction efficiency for $\mu^{+}$ at ICAL}
 \end{figure}

Reconstruction efficiency ($\varepsilon_{rec}$) for each  $\mu (E_{\mu},\cos\Theta)$ event is given by the ratio of total number of events properly reconstructed  
($\eta_{rec}$) and total number of events generated in the central region of the detector ($\eta_{total}$) i.e. $\varepsilon_{rec} = \frac{\eta_{rec}}{\eta_{total}}$. Fig.~\ref{fig:16} shows the reconstruction efficiencies of the muon events as a function of $E_{\mu}$ for various $\cos\theta$ values. The left and right figures are for $\mu^{-}$ and $\mu^{+}$ respectively. We can see from the figure how the reconstruction efficiency depends on the true energy and true zenith angle of the muon. The efficiency is seen to initially rise with muon energy, reach a peak at about $E_\mu \sim 10$ GeV, after which it is seen to fall albeit extremely slowly. The dependence on $\cos\theta$ is more complicated. For lower energies the muon reconstruction is better for upward going neutrinos, with the nearly horizontal muons not being reconstructed at all since the these muons fail to cross even 3 RPCs since the RPCs are arranged horizontally.  However, as the muon energy increase beyond $E_\mu > 20 $ GeV, the length of the muon track increases and thereafter it becomes easier for the more horizontal muons to be reconstructed due to the rectangular geometry of ICAL.\\

\begin{figure}[tbp]
   \centering
     \includegraphics[width=.45\textwidth]{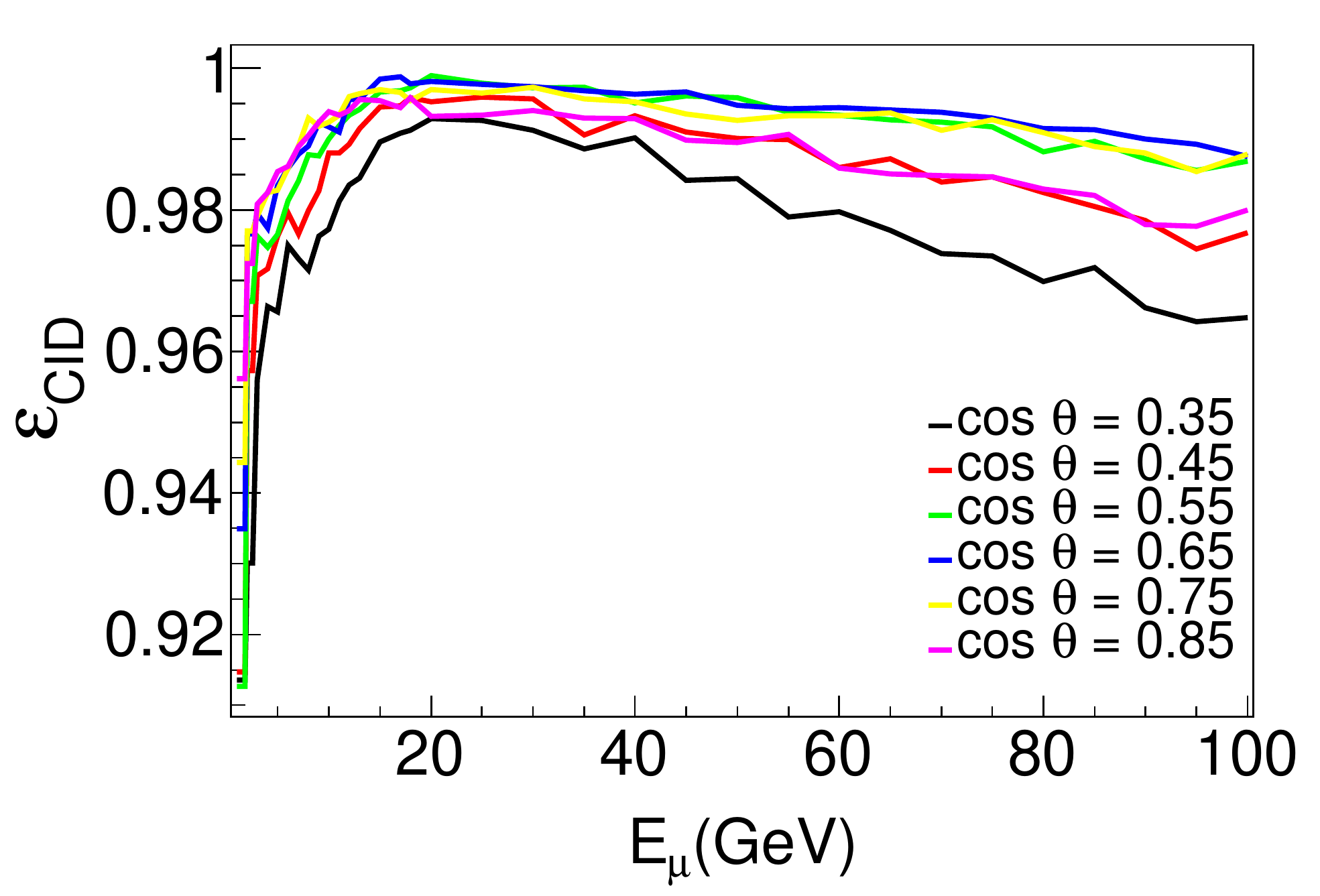}
     \hfill
     \includegraphics[width=.45\textwidth]{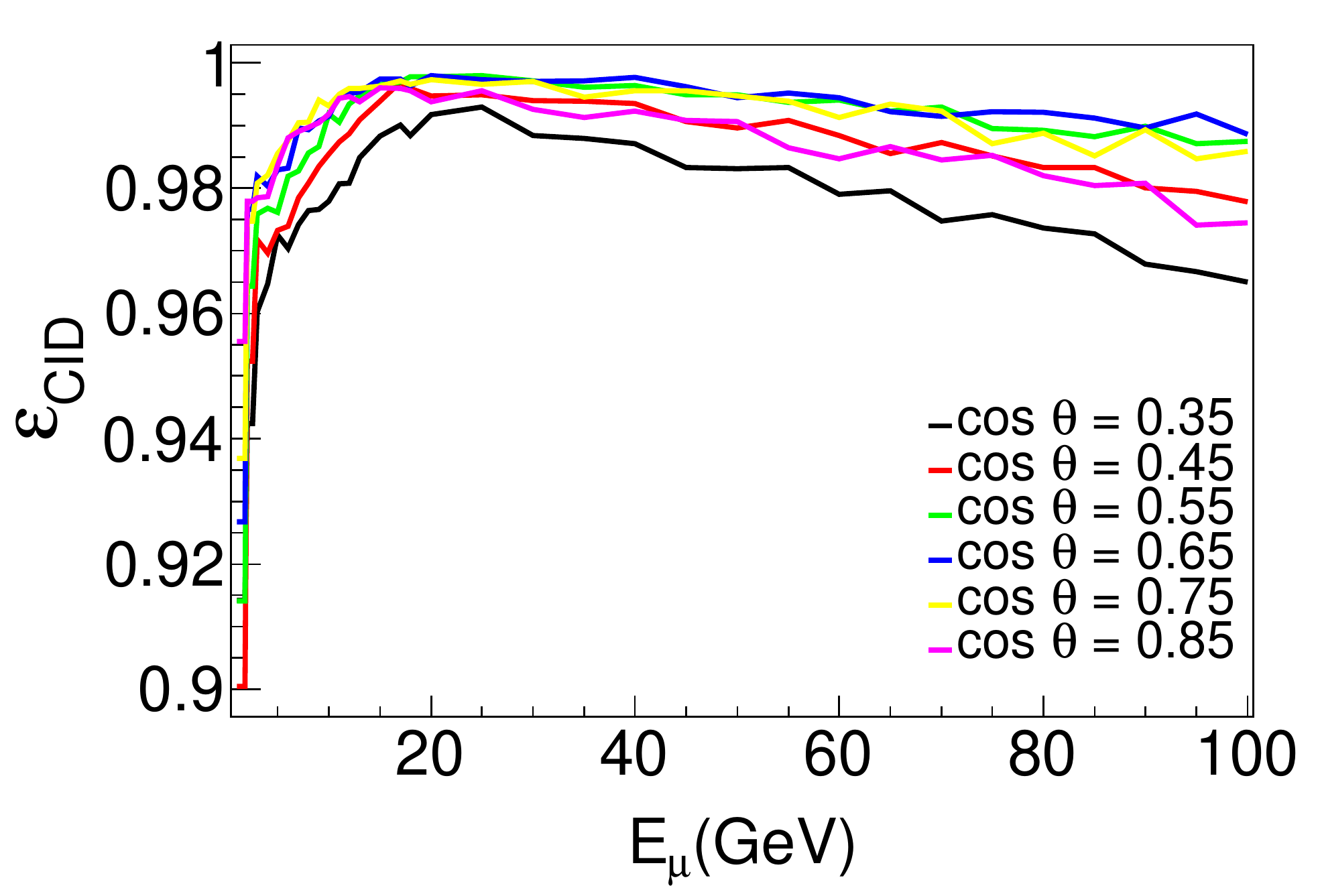}
     \caption{\label{fig:17} LEFT: Charge identification reconstruction efficiency for $\mu^{-}$ at ICAL RIGHT: Charge identification reconstruction efficiency for $\mu^{+}$ at ICAL}
 \end{figure}

Charge Identification Efficiency($\varepsilon_{CID}$) is defined as the ratio of the events with correctly identified charges($\eta_{CID}$) to the total reconstructed ($\eta_{rec}$) events: $\varepsilon_{CID}=\frac{\eta_{CID}}{\eta_{rec}}$. Fig.~\ref{fig:17} shows the CID reconstruction efficiencies of the muon events as a function of $E_{\mu}$ for various $\cos\theta$. Again, the left and right figures are for $\mu^{-}$ and $\mu^{+}$ respectively. The charge identification efficiency is seen increase until $E_\mu \sim 20$ GeV and thereafter fall. The dependence on the muon zenith angle is again seen to be complicated. However, Fig. ~\ref{fig:17}  reveals that in the energy region of our interest, {\it i.e.}, $E_\mu = (1-100)$ GeV, the charge identification efficiency in ICAL is better than 96\% for all muon zenith angles. 

\begin{figure}[tbp]
   \centering
     \includegraphics[width=.45\textwidth]{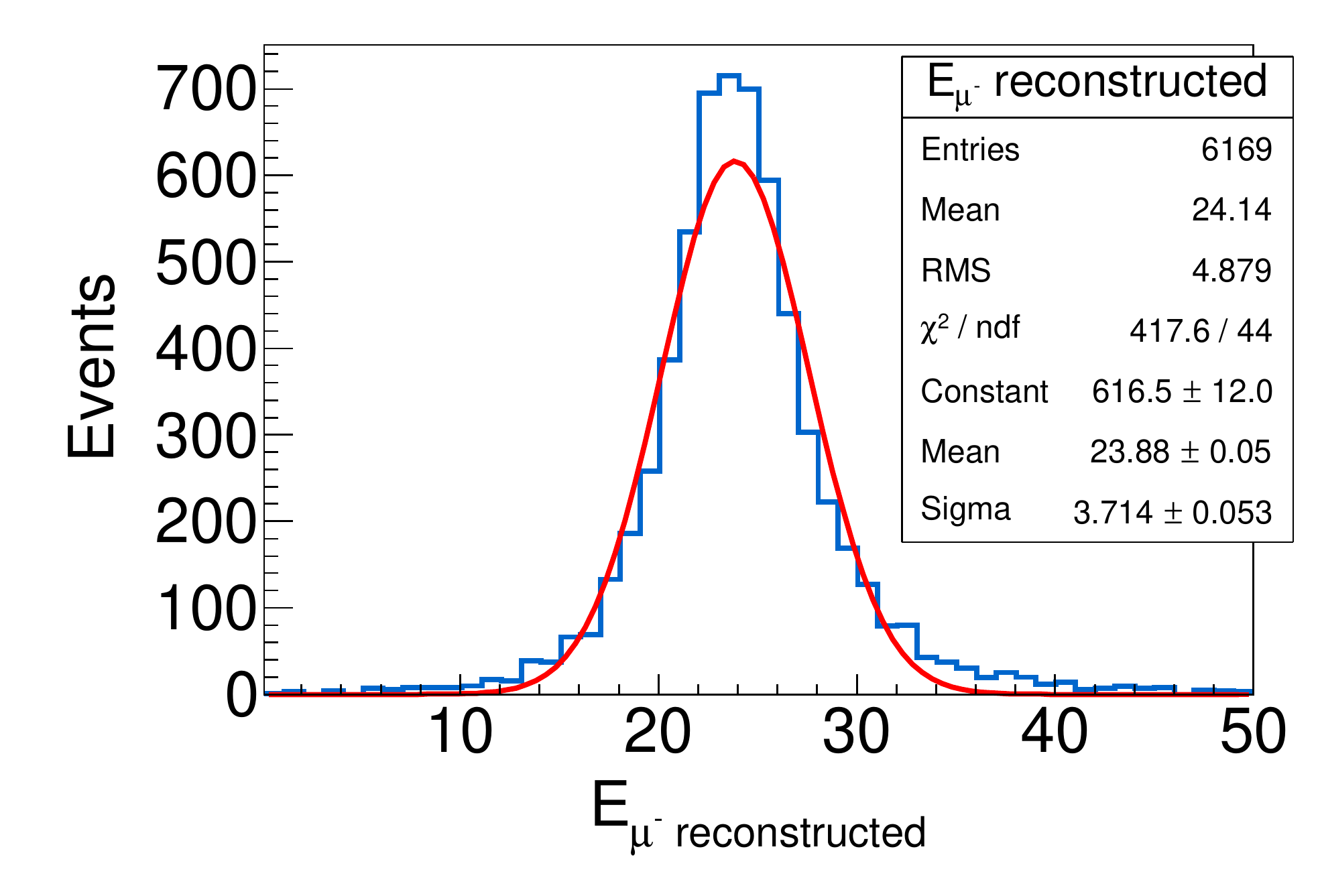}
     \hfill
     \includegraphics[width=.45\textwidth]{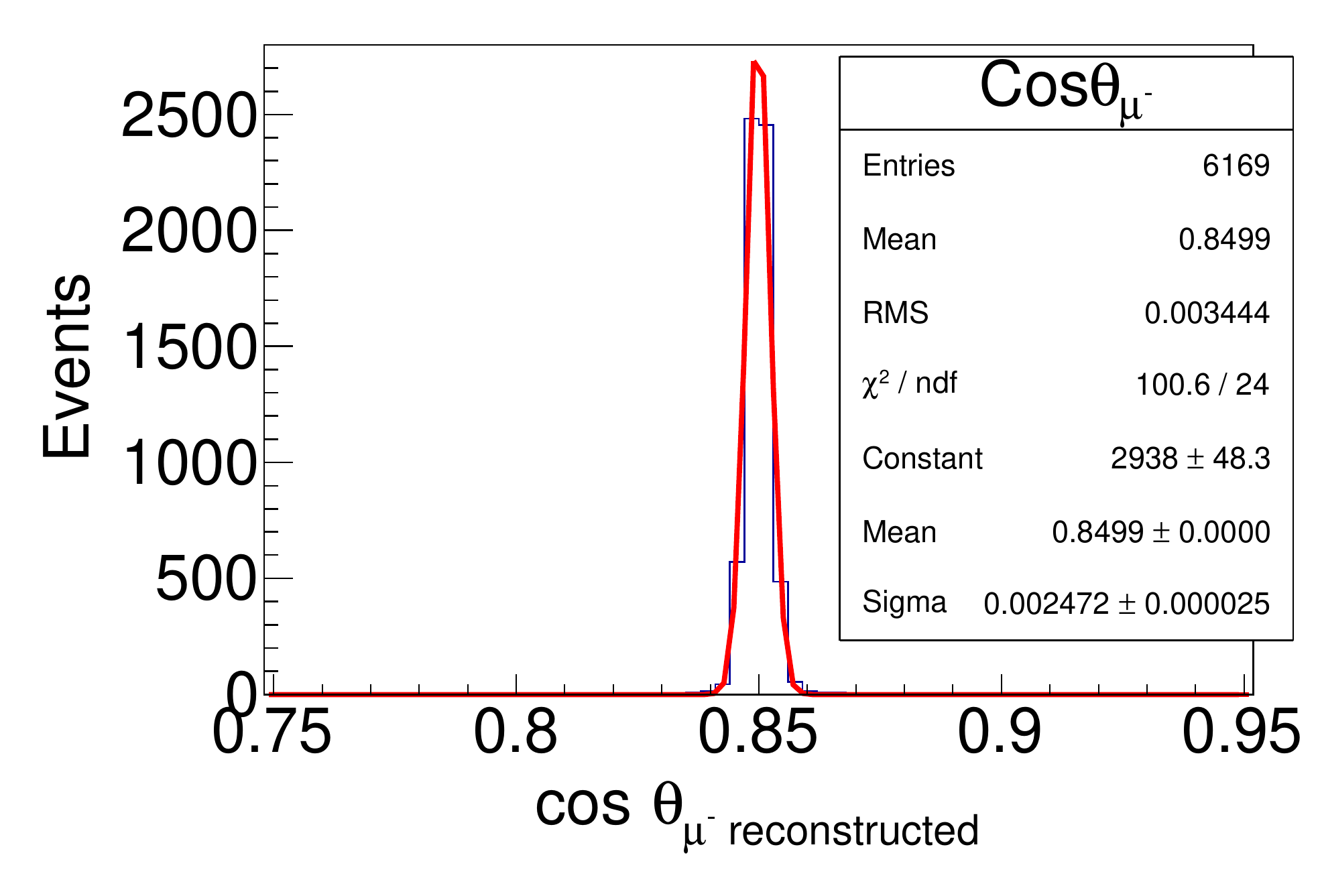}
     \caption{\label{fig:18} LEFT: Reconstructed momentum distribution for $\mu^{-}$ at ICAL. The energy resolution($\sigma_{E}$) is given by $\sigma_{E}^{\prime}/E$, where  $\sigma_{E}^{\prime}$ is the width obtained by fitting this with Gaussian probability distribution functions. RIGHT: Reconstructed cosine of zenith angle distribution for $\mu^{-}$ at ICAL. Angular resolution is given by $\sigma_{cos\theta}$ which is the width obtained by fitting it with Gaussian probability distribution functions. Both the distributions are for $\mu^{-}$ with true $E_{\mu} = 25$ GeV and $\cos\theta = 0.85$.}
 \end{figure}

\begin{figure}[tbp]
   \centering
     \includegraphics[width=.45\textwidth]{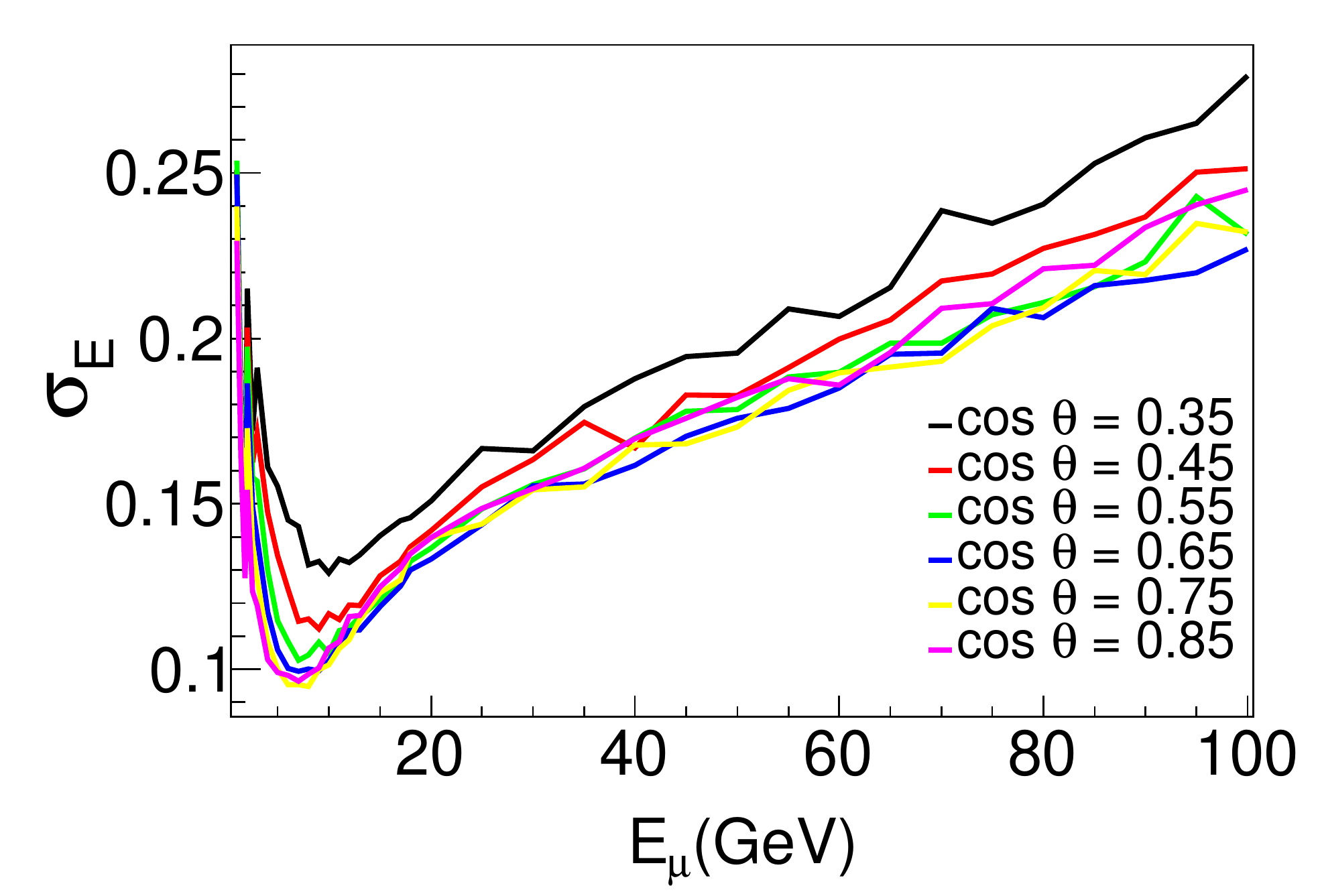}
     \hfill
     \includegraphics[width=.45\textwidth]{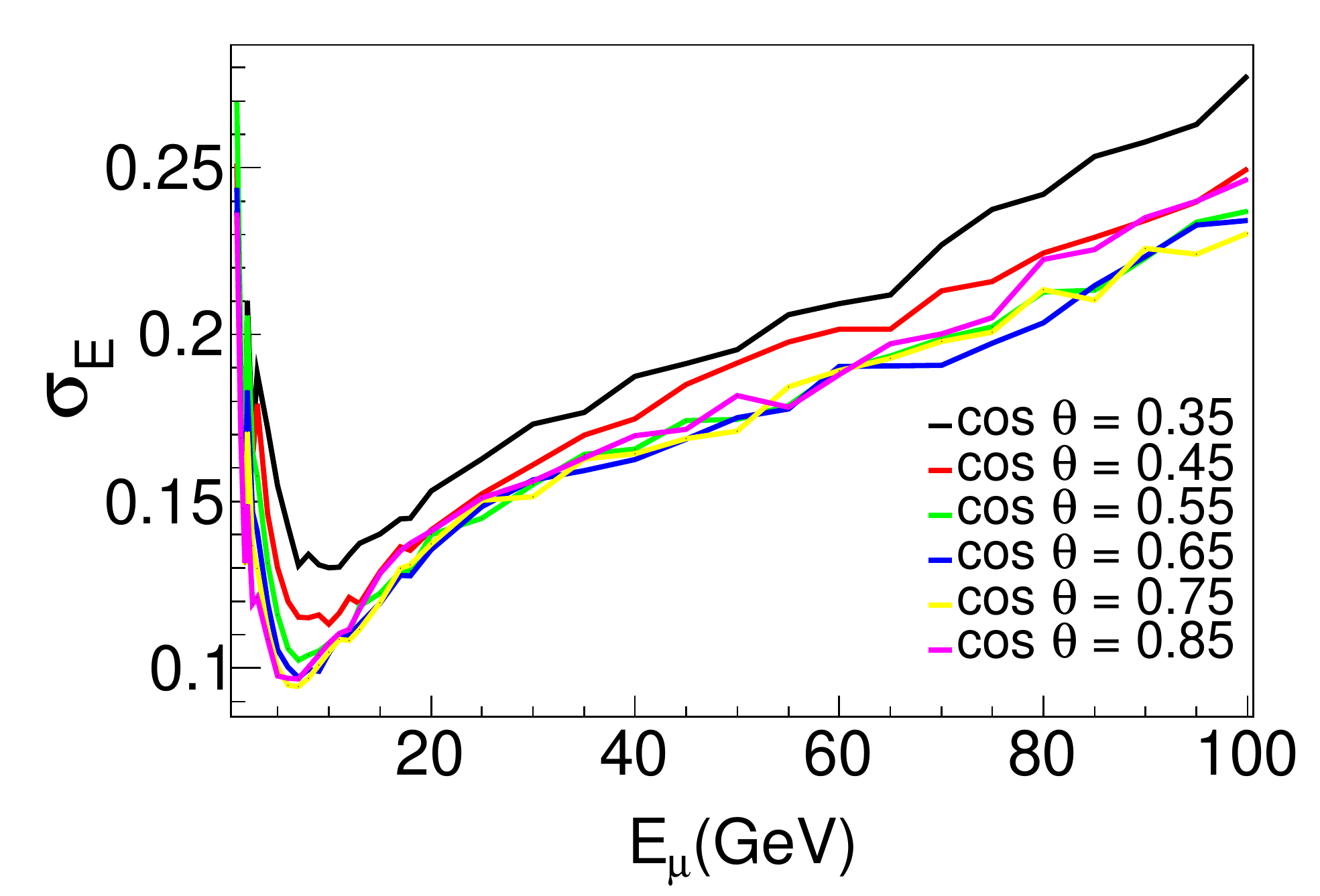}
     \caption{\label{fig:19} LEFT: Momentum resolution for $\mu^{-}$. RIGHT: Momentum resolution for $\mu^{+}$ at ICAL}
 \end{figure}

In order to obtain the muon energy resolutions, the reconstructed momentum distributions for $\mu^-$ (and $\mu^+$) are plotted as a function of $E_{\mu}$ for a given true $E_\mu$ and true muon $\cos\theta$ value, and then fitted with a Gaussian function to get the $\sigma_E$ of the distribution. The left panel of Fig.~\ref{fig:18} shows the reconstructed momentum distribution for $\mu^{-}$ for true muon $E_\mu =25$ GeV and true muon $\cos\theta = 0.85$. The fitted value of $\sigma_E$ can be read off from the figure. This process is repeated for all values of the true muon energy and true muon zenith angle. The left panel of Fig.~\ref{fig:19} shows the $\sigma_E$ for a $\mu^{-}$ as a function of true muon energy $E_\mu$ and for the full set of values of $\cos\theta$, while the right panel shows the corresponding plots for $\mu^+$. \\

\begin{figure}[tbp]
   \centering
     \includegraphics[width=.45\textwidth]{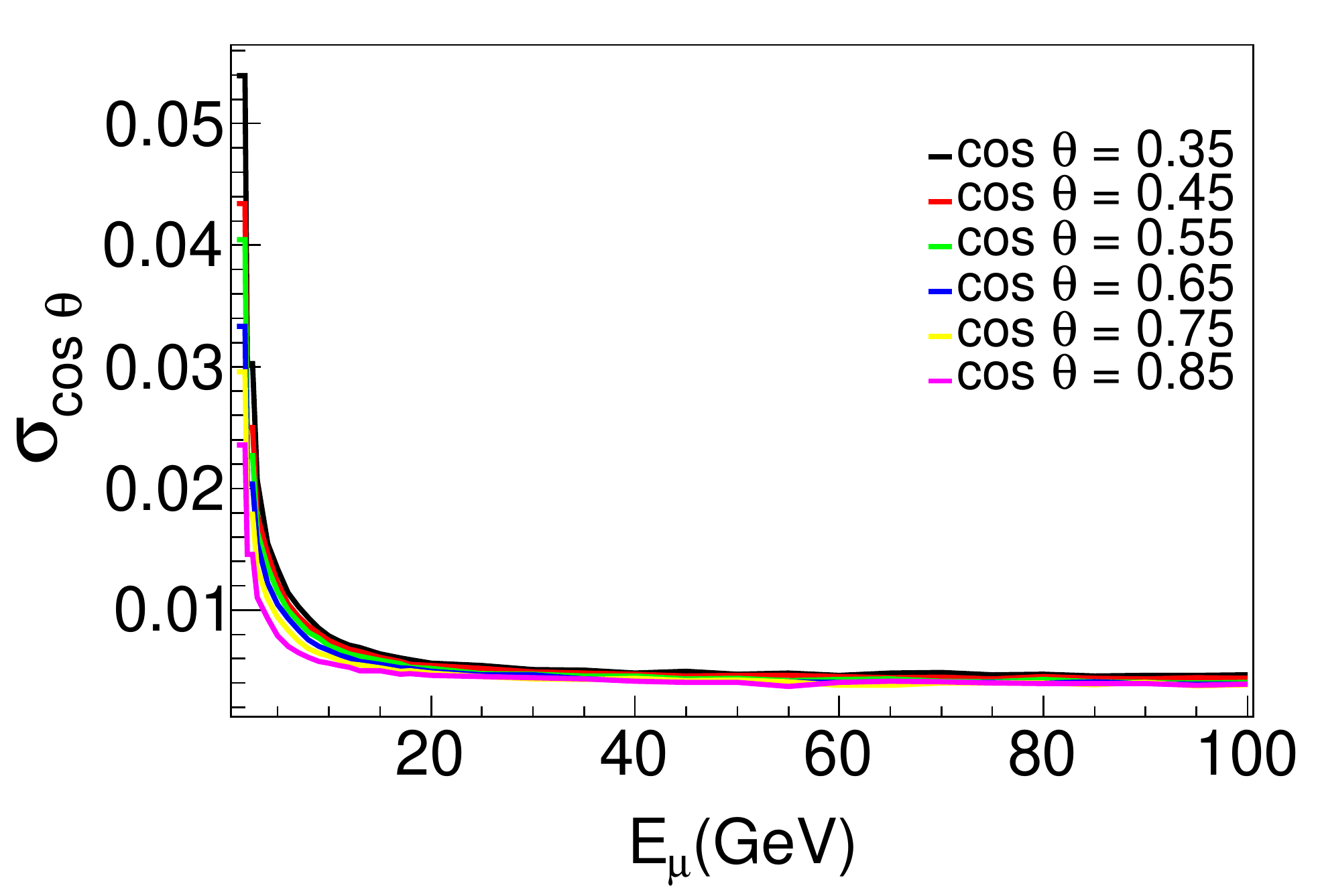}
     \hfill
     \includegraphics[width=.45\textwidth]{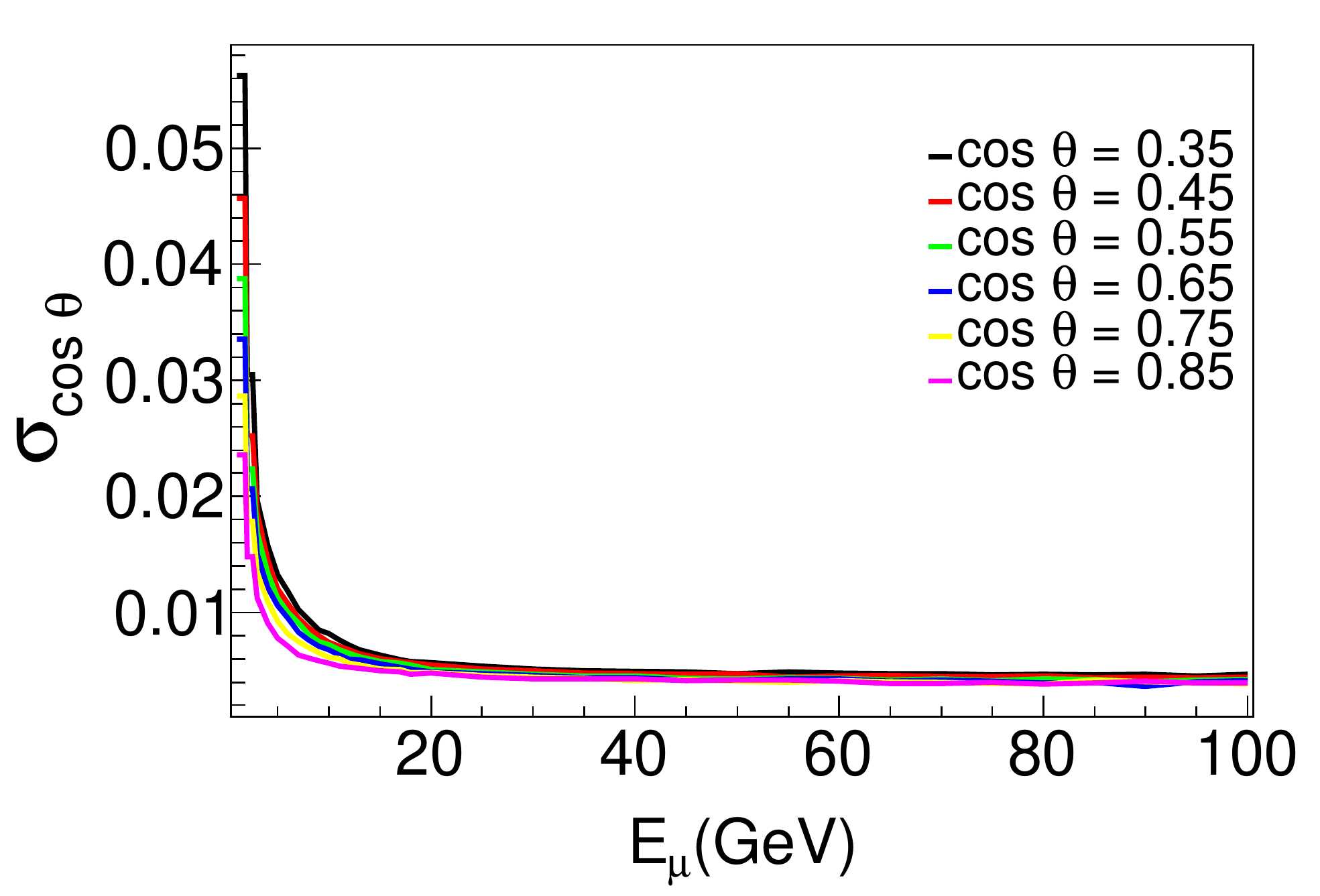}
     \caption{\label{fig:20} LEFT:  $\cos\theta$ resolution for $\mu^{-}$ at ICAL. RIGHT: $\cos\theta$ resolution for $\mu^{+}$ at ICAL}
 \end{figure}

To obtain the muon zenith angle resolution we use a similar procedure. The right panel of Fig.~\ref{fig:18} shows the reconstructed zenith angle distribution for $\mu^{-}$ for true muon $E_\mu =25$ GeV and true muon $\cos\theta = 0.85$. The width of the distribution gives $\sigma_{\cos\theta}$ which is extracted from the fit and the steps repeated for all values of true muon energy and true muon zenith angle to get the full table. The left panel of Fig.~\ref{fig:19} shows the $\sigma_E$ for a $\mu^{-}$ as a function of true muon energy $E_\mu$ and for the full set of values of $\cos\theta$, while the right panel shows the corresponding plots for $\mu^+$. Similarly, the left panel of Fig.~\ref{fig:20} shows the $\sigma_{\cos\theta}$ for a $\mu^{-}$ as a function of true muon energy $E_\mu$ and for the full set of values of $\cos\theta$, while the right panel shows the corresponding plots for $\mu^+$.

\bibliographystyle{JHEP}
\bibliography{ref}
\end{document}